\documentclass[aps, prd, reprint, nofootinbib]{revtex4-2}
\usepackage{graphicx}
\usepackage{amsmath,amssymb}
\usepackage{dcolumn}
\usepackage{bm}
\usepackage{xcolor}
\usepackage{soul}
\usepackage{tikz}
\usepackage{subcaption}
\usepackage{comment}
\usepackage{ragged2e}
\usepackage{caption}
\usepackage[linktocpage=true]{hyperref}
\hypersetup{
colorlinks=true,
citecolor=darkblue,
linkcolor=reddish,
urlcolor=darkblue,
pdfauthor={},
pdftitle={},
pdfsubject={}
}
\newcommand{\mnras}{MNRAS}

\definecolor{blue2}{cmyk}{1, 0.1, 0.1, 0.1}

\usepackage{colortbl}
\definecolor{lightgreen}{cmyk}{0.2, 0, 0.2, 0.2}
\definecolor{lightgray2}{cmyk}{0.1,0.1,0,0.1}
\definecolor{Red2}{RGB}{214, 39, 40}
\definecolor{Blue2}{RGB} {31, 119, 180}
\definecolor{Orange2}{RGB}{255, 127, 14}
\definecolor{Green2}{RGB}{44, 160, 44}
\definecolor{greyish2}{rgb}{.96,.96,.96}

\definecolor{Red}{RGB}{214, 39, 40}
\definecolor{Blue}{RGB} {31, 119, 180}
\definecolor{Orange}{RGB}{255, 153, 51}
\definecolor{Purple}{RGB}{178, 102, 255}
\definecolor{Green}{RGB}{44, 160, 44}
\definecolor{regal}{RGB}{90,0,120}
\definecolor{darkblue}{rgb}{0.15,0.35,0.55}
\definecolor{reddish}{rgb}{0.65, 0.2, 0.2}
\definecolor{darkgreen}{RGB}{50,150,0}
\definecolor{greyish}{rgb}{.90,.90,.90}
\definecolor{greyish2}{rgb}{.96,.96,.96}
\definecolor{greyish3}{rgb}{.37,.37,.37}
\definecolor{darkblue2}{rgb}{0.3,0.4,0.9}
\definecolor{Blue3}{RGB}{31, 119, 180}

\definecolor{pyBlue}{RGB}{31, 119, 180}
\definecolor{pyRed}{RGB}{214, 39, 40}
\definecolor{pyGreen}{RGB}{44, 160, 44}
\definecolor{pyBlue2}{RGB}{0, 111, 237}
\definecolor{pyRed2}{RGB}{224, 52, 36}
\definecolor{Mathematica1}{rgb}{0.368417, 0.506779, 0.709798}
\definecolor{Mathematica2}{rgb}{0.880722, 0.611041, 0.142051}

\def\beq{\begin{equation}}
\def\eeq{\end{equation}}

\usetikzlibrary{shapes.misc}

\tikzset{cross/.style={cross out, draw=black, minimum size=2*(#1-\pgflinewidth), inner sep=0pt, outer sep=0pt},
cross/.default={1pt}}

\graphicspath{ {./figures} }

\begin{document}

\title{Simulating Binary Primordial Black
Hole Mergers in Dark
Matter Halos}

\author{Muhsin Aljaf}
\email{muhsinaljaf@oakland.edu, orcid.org/0000-0002-1251-4928}
\affiliation{Department of Physics, Oakland University, Rochester, Michigan, 48309, USA}
\author{Ilias Cholis}%
\email{cholis@oakland.edu, orcid.org/0000-0002-3805-6478}
\affiliation{Department of Physics, Oakland University, Rochester, Michigan, 48309, USA}

\date{\today}

\begin{abstract}
Primordial black holes (PBHs), possibly constituting a non-negligible fraction of dark matter (DM), might be responsible for a number of gravitational wave events detected by LIGO/Virgo/KAGRA. In this paper, we simulate the evolution of PBH binaries in DM halos and calculate their merger rate up to redshift of 10. We assume that DM halos are made entirely by a combination of single PBHs and PBH binaries. 
We present the resulting merger rates from  the two main channels that lead to merging PBH binaries: two-body captures and binary-single interactions. We account for alternative assumptions on the dark matter halo mass-concentration relationship versus redshift. We also study what impact the PBH mass distribution, centered in the stellar-mass range, has on the PBH merger rate that the ground-based gravitational-wave observatories can probe. 
We find that under reasonable assumptions on the abundance of PBH binaries relative to single PBHs, the binary-single interaction rates can be dominant over the two-body capture channel.  Our work studies in detail the dynamics of PBHs inside DM halos, advancing our understanding on how the current
gravitational-wave events constrain the properties of PBHs. Moreover, we make predictions in a redshift range to be probed by future observatories. 
\end{abstract}

\maketitle


\section{Introduction}\label{sec:intro}

Primordial black holes (PBHs) are formed from primordial perturbations in the early universe through a variety of mechanisms~\cite{PBH_1, PBH_2, PBH_3, PBH_4}. Following the discovery of the first gravitational wave event in 2015 \cite{LIGOScientific:2016aoc}, interest in PBHs grew significantly. Depending on their mass, PBHs could contribute some fraction or even all dark matter (DM) and be detectable in gravitational waves \cite{Bird:2016dcv, PBH_DM_1,PBH_DM_2,PBH_DM_3, Kovetz:2016kpi, Clesse:2020ghq, Clesse:2018ogk}. 
Conversely, gravitational wave observations can also be used to set limits on the abundance of PBHs \cite{early_PBH_1, PBBH1, Kovetz:2017rvv, Morras:2023jvb, LIGOScientific:2022hai}.
In addition, PBHs might serve as seeds for the formation of supermassive black holes\cite{PBH_super_1,PBH_super_2,PBH_super_3,PBH_super_4}.

The binary formation mechanisms of PBH vary on the basis of their formation time and interaction type, resulting in distinct properties (see e.g. Ref.~\cite{PBH_channels}). In the late universe, binaries form within DM halos via dynamical capture or binary-single interactions. In dynamical capture, PBH binaries emerge from hyperbolic encounters that lose sufficient energy through gravitational wave emission. The three-body channel involves $2+1$ PBH encounters, where the third PBH carries away enough energy for the other two to become bound. 
The merger rate of PBH binaries via direct capture is estimated in \cite{Bird:2016dcv, Mandic}, showing its dominance in small mass halos, with resulting binaries being highly eccentric and merging quickly \cite{Cholis:2016kqi}. The merger rate from the single binary channel within the PBH mini-halos, examined in \cite{Kritos}, predominates over the direct capture rate only when the fraction of PBH in DM is enhanced as such interactions are likely to occur in dense environments.

On the other hand, in the early universe, PBH binaries can form when nearby PBH pairs are decoupled from the Hubble flow~\cite{early_PBH_1,PBBH1}. Ref.{\cite{PBBH1}} 
showed that such binaries are highly eccentric. In addition, it calculated the rate at which such binaries merge today and showed that other than a small fraction, binaries are not perturbed by PBH tidal torques and encounters with other PBHs in their environment. However, their work relied on analytical calculations valid for massive halos and in regions of the halos far from their centers. 

In this work, we evaluate the PBH merger rate due to direct capture events and due to the evolution of the orbital properties of PBH binaries from interactions with other (single) PBHs, inside DM halos.  We focus on stellar-mass PBHs that can be directly probed by the LIGO/Virgo/KAGRA ongoing gravitational wave observations.
We account fully for the mass-distribution of dark matter halos, also known as the halo mass function, and how that evolves with redshift. We perform calculations including DM halos with current masses from $10^{3}$ to $10^{15} \, M_{\odot}$. 
We simulate the entire volume of dark matter halos from the redshift of 12 to the present era. This is important to take into account, as other than the smallest mass $O(10^3-10^{4}) M_{\odot}$ halos, there is a significant density gradient inside DM halos and also a gradient in the velocity distribution of PBHs. This can affect the derived PBH merger rates. Our dark matter halos' total mass, PBH density profiles (described by the concentration parameter) and PBH velocity distribution profiles evolve with time as well.  

We describe all our assumptions regarding the modeling of the DM halos and how those evolve in Section~\ref{sec:DM_haio_models}.
In Section~\ref{sec:Two_body_captures}, we describe our methodology in evaluating the PBH-PBH direct capture interactions and our subsequent results.
We study both the simple monochromatic $30 \, M_{\odot}$ case, which has been studied in the past (as e.g. in \cite{Bird:2016dcv}), but also more realistic PBH mass distributions with probability density functions peaking approximately at the same mass. Our results are in agreement with \cite{Bird:2016dcv, Mandic}, at low redshifts and by a factor of $\sim 2$ at a redshift of 6, increasing rapidly at earlier times.
These findings are robust to the PBH mass distribution and DM halo concentration relation. 
In Section~\ref{three_body}, we discuss the binary PBH-single PBH interactions occurring inside DM halos. We study these interactions for the entire DM halo masses range, simulating the evolution of these binaries under different environmental (PBH density and velocity distribution) conditions. These interactions accelerate the PBH binary evolution, especially in the inner parts of the DM halos. We find that if an appreciable fraction of PBHs are in binaries at formation, then the PBH merger rate is enhanced compared to just the direct capture calculation. 
Finally, in Section~\ref{sec:conclusions}, we combine all our results and discuss further connections to future gravitational wave observations.

\section {DARK MATTER HALO MODELS}
\label{sec:DM_haio_models}

In this section, we briefly discuss the DM halo models used in this paper. This includes key quantities such as the density profile, concentration parameter, and halo mass function, which are important for the calculation of the merger rate of PBH binaries from a) two-body captures and b) binary-single interactions.

\subsection{The Halo Profile}
We take the PBH density profile to follow the  Navarro, Frenk, and White (NFW) DM profile \cite{Navarro:1995iw}, 
\begin{equation}   
\rho_{\textrm{NFW}}(r)=\frac{\rho_s}{\left(r / R_{s}\right)\left(1+r / R_{s}\right)^2}.
\end{equation}
Parameter $r$ is the distance from the halo's center, while $R_{s}$ is the scale radius of the halo. The DM density $\rho_{s}$ is defined as $\rho_{\textrm{crit}} \cdot \delta_{c}$, where $\rho_{\textrm{crit}}$ denotes the critical density of the Universe at a specific redshift $z$, and $\delta_{c}$ is the linear overdensity threshold. The overdensity threshold $\delta_{c}$ is connected to the concentration parameter $C$ through,
\begin{equation}   
\delta_{c}=\frac{200}{3} \frac{C^3}{g(C)},
\end{equation}
where 
\begin{equation}
g(C)=\ln (1+C)-\frac{C}{1+C}.
\end{equation}

The concentration parameter $C\equiv \frac{R_{\textrm{vir}}}{R_s}$, characterizes the
central density of DM halos. The halo's virial radius, $R_{\textrm{vir}}$ covers a volume
within which the average halo density is 200 the critical density of the Universe.

Given a the halo density profile, we can calculate the total mass within a sphere of radius $R_{\textrm{vir}}$ as,
\begin{eqnarray}
M\left(R_{\textrm{vir}}\right)&=&4 \pi \int_0^{R_{\textrm {vir }}}d r r^2 \rho_{\textrm{NFW}}(r)
\nonumber \\
&=&4 \pi R_s^3 \rho_s\left[\ln \frac{R_{\textrm{vir}}+R_s}{R_s}-\frac{R_{\textrm{vir}}}{R_{\textrm{vir}}+R_s}\right]
\\
&=&4 \pi R_s^3 \rho_s\left[\ln (1+C)-\frac{C}{1+C}\right]=4 \pi R_s^3 \rho_s g(C). \nonumber
\label{rh0_s} 
\end{eqnarray}

We note that the mass of the halo is a function of the concentration. In the above equation, $g(C)$ is just the quantity in square brackets.

\subsection{The Mass-Concentration-Redshift Relation $C(M,z)$}

As we mentioned above, the concentration parameter plays an important role in the properties of DM halos. $N$-body simulations show that the concentration parameter decreases with increasing halo mass and varies with redshift at a fixed mass~\cite{Prada12,Ludlow:2016ifl}.
This behavior is consistent with the dynamics of the merger tree of DM halos and their evolution, where smaller halos, having already virialized, tend to be more concentrated than larger ones. 
The relationship between concentration, mass, and redshift can be described by the $C(\nu)$ relation. Here, $\nu(M,z)$, the peak height, is a dimensionless parameter defined as $\nu(M,z) \equiv \delta_{\textrm{sc}}(z)/\sigma(M, z)$, where $\delta_{\textrm{sc}}(z)=1.686(1+z)$ is the spherical collapse threshold for overdensities and $\sigma(M, z)$ is the linear root mean square fluctuation of overdensities. This peak height parameter indicates that the concentration parameter depends on both the mass and the redshift.

\begin{figure}[ht!]
    \centering
\includegraphics[width=\linewidth]{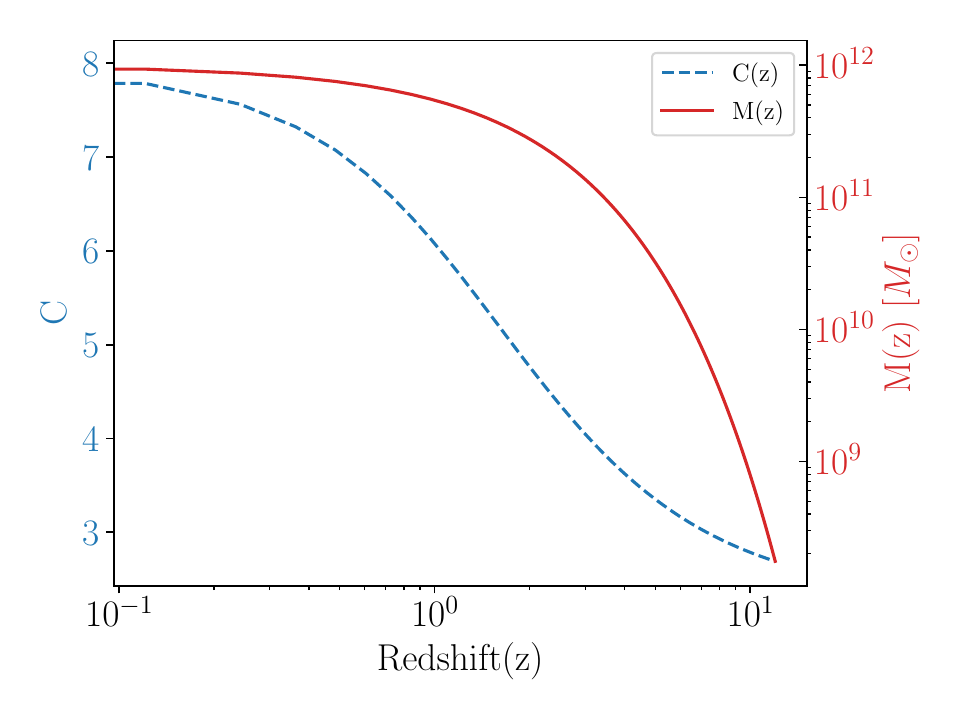}
    \caption{\justifying The redshift evolution of mass and concentration parameter $C$  for a $10^{12} M_\odot$ halo at z=0, using \textit{Ludlow16} model. 
    }
    \label{fig:CONC_MASS_Z}
\end{figure}

In this paper, we utilize two key models for the mass-concentration-redshift relation $C(M,z)$ that give a good fit to the DM N-body simulation results.
Henceforth, we shall refer to these models as \textit{Ludlow16} \cite{Ludlow:2016ifl} and \textit{Prada12} \cite{Prada12}. 
For \textit{Ludlow16}, we employ Eq.~C1 of Ref.~\cite{Ludlow:2016ifl}. For \textit{Prada12}, we use the model described by equations Eqs.~12-22 of Ref.~\cite{Prada12}. In Fig~\ref{fig:CONC_MASS_Z}, for a DM halo that reaches a mass of $10^{12} \, M_{\odot}$ at $z=0$; we show how its mass and concentration parameter evolved with redshift starting from $z=12$. We used the \textit{Ludlow16}
concentration evolution model.

\begin{figure}[htb!]
   \centering
   \includegraphics[width=\linewidth] {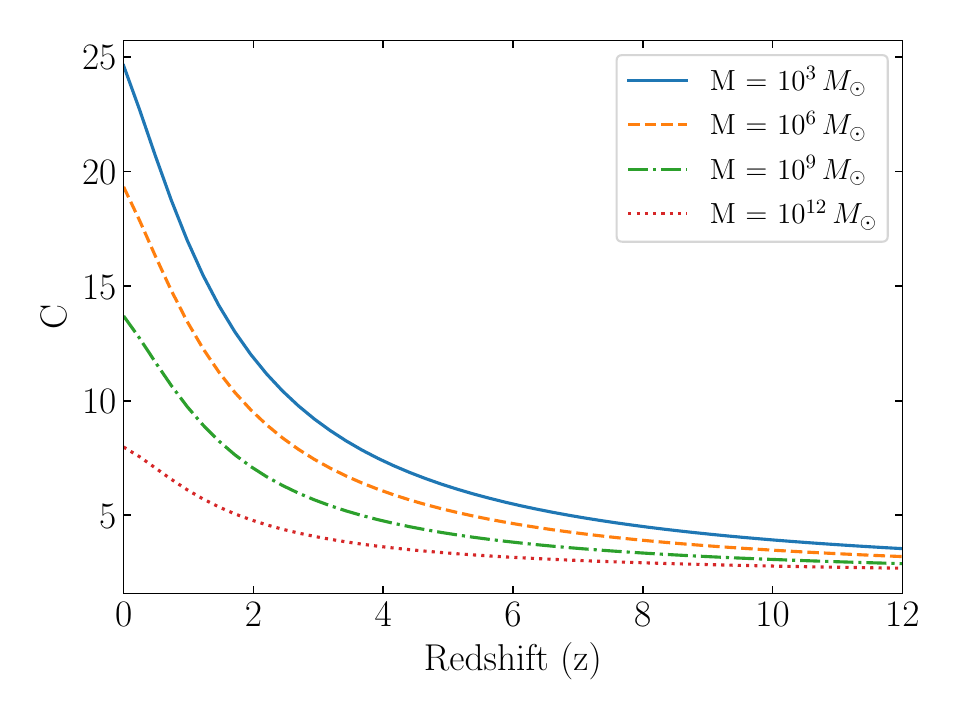} \\
   \vspace{-0.2cm}
   \includegraphics[width=\linewidth]{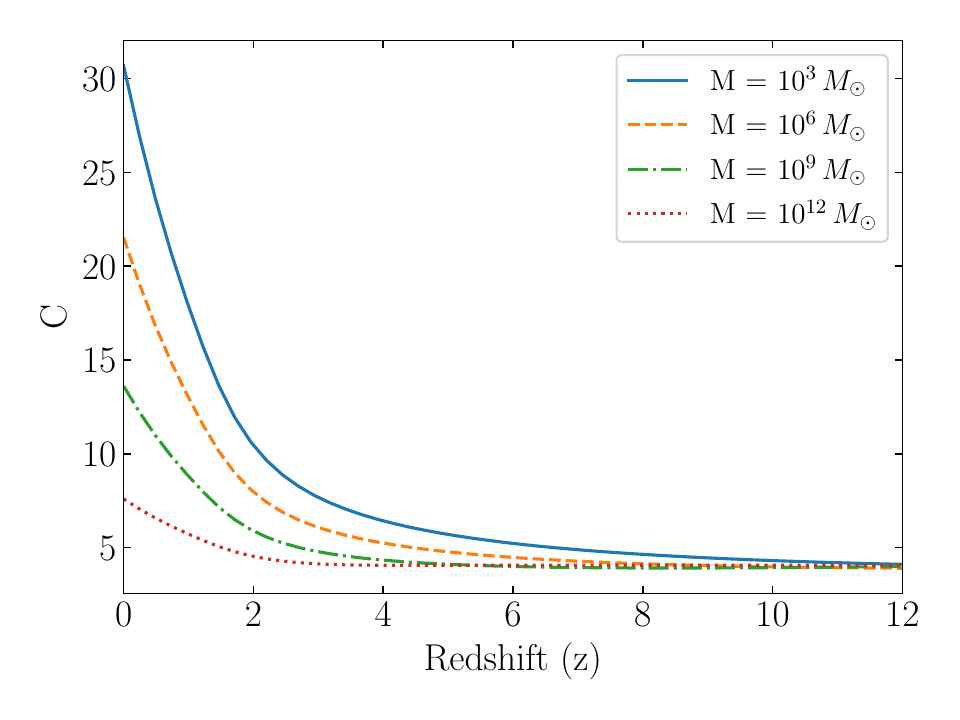}
   \vspace{-0.7cm}
   \caption{\justifying
   The redshift evolution of the concentration parameter for various DM halos. The expectations of the \textit{Ludlow16} (\textit{top}) and \textit{Prada12} (\textit{bottom}) concentration models are shown for halos that at present ($z=0$) have masses of $10^3, 10^6, 10^9$ and $10^{12}$ $M_{\odot}$.}
    \label{fig:Conc}
\end{figure}

In Fig.~\ref{fig:Conc}, we illustrate the concentration parameter of DM halos of fixed mass, for the \textit{Ludlow16} model (top panel)  and for the \textit{Prada12} model (bottom panel). We note that in our calculations we evolve the mass of DM halos from accretion and use such relations to find what the concentration parameter is for a halo of a given mass at a given redshift. 

\subsection{The Halo Mass Function}

A key quantity in determining the merger rate of PBH binaries is the halo mass function.
This describes the mass-distribution of DM halos. 
We use the differential halo mass function as introduced in \cite{PS},
\begin{equation}
\frac{d n}{d \ln M}=M \cdot \frac{\rho_{m_{0}}}{M^2} f(x)\left|\frac{d \ln x}{d \ln M}\right|.
\end{equation}
$n$ is the number density of DM halos, $M$ is the halo mass, $\rho_{m_{0}}$ is the mean density of matter, and $f(x)$ is a function related to the geometrical conditions for the overdensities at the collapse time of the halo. $f(x)$ can be derived from analytical work or numerical simulations.
We use the halo mass function by \textit{Press-Schechter} \cite{PS},
\begin{equation}
f(x)=\sqrt{\frac{2}{\pi}} \frac{\delta_c}{x} \exp \left(-\frac{\delta_c^2}{2 x^2}\right).
\end{equation} 

The DM mass function is readily accessible through the publicly available Python package \texttt{HMFcalc} \cite{HMFcalc}. To estimate the total merger rate due to capture and binary-single interactions of all DM halos,
we will first evaluate the merger rate per DM halo $R_{\textrm{halo}}$ of mass $M$.
Then we will integrate that rate over the halo mass function.

\subsection{Primordial Black Hole Velocity Distributions}
To study the interactions between PBHs, let's consider two PBHs with masses $m_1$ and $m_2$ and a relative velocity $v_{\textrm{rel}} = |v_1 - v_2| = v_{\textrm{pbh}}$. 
Those PBHs are in a DM halo with virial velocity $v_{\textrm{vir}}$ and dispersion velocity $v_{\textrm{disp}}$. 
The relative velocity distribution of PBHs can be approximated by a truncated Maxwell-Boltzmann distribution~\cite{Bird:2016dcv, Cholis:2016kqi},
\begin{equation}
p(v_{pbh}) = F_0^{-1} v_{\textrm{pbh}}^{2}\left[\exp\left(-\frac{v_{\textrm{pbh}}^2}{v_{\textrm{disp}}^2}\right) 
- \exp\left(-\frac{v_{\textrm{vir}}^2}{v_{\textrm{disp}}^2}\right)\right]. 
\end{equation}
$F_0$ is a normalization constant such that,

\begin{equation}
F_0 = 4\pi \int_0^{v_{\textrm{vir}}} v_{pbh}^2 \left( e^{-v_{pbh}^2/v_{\textrm{disp}}^2} - e^{-v_{\textrm{vir}}^2/v_{\textrm{disp}}^2} \right) \, dv.
\end{equation}

The relationship between $v_{\textrm{disp}}$ and $v_{\textrm{vir}}$ can be derived from the properties of the halo as, 
\begin{eqnarray}
v_{\textrm{disp}} &=& \sqrt{\frac{G M(r<r_{\textrm{max}})}{r_{\textrm{max}}}} \nonumber \\
&=& \frac{v_{\textrm{vir}}}{\sqrt{2}} \sqrt{\frac{C}{x_{\textrm{max}}} \frac{g(x_{\textrm{max}})}{g(C)}},
\label{v_disp}
\end{eqnarray}
where $x_{\textrm{max}}  = \frac{r_{\textrm{max}}}{R_s} = 2.1626$ ($C_m$ of Ref.~\cite{Bird:2016dcv}) and
\begin{eqnarray}
M(r < r_{\textrm{max}}) &=& 4 \pi \int_0^{r_{\textrm{max}}} dr \, r^2 \, \rho_{\textrm{NFW}}(r) \nonumber \\
&=& 4 \pi R_s^3 \rho_s \int_0^{x_{\textrm{max}}} dx \, \frac{x}{(1 + x)^2} \\
&=& 4 \pi R_s^3 \rho_s \left[\ln(1 + x_{\textrm{max}}) - \frac{x_{\textrm{max}}}{1 + x_{\textrm{max}}}\right]. \nonumber
\end{eqnarray}
Given $R_{\textrm{s}}, \rho_{\textrm{s}}$ and $C$, we calculate the velocity dispersion $v_{\textrm{disp}}$ from the first equality of Eq.(\ref{v_disp}) and the virial velocity $v_{\textrm{vir}}$ from the second equality.

\section{Two-body capture events}
\label{sec:Two_body_captures}

In this section, we describe our assumptions for the evaluation of the two-body capture rate between PBHs, following Ref.~\cite{Bird:2016dcv}. Since the binaries formed from such interactions are hard binaries with high initial eccentricities and small semi-major axes (see Ref.~\cite{Cholis:2016kqi} for more details), we also consider that the merger of the binaries follows with a small delay in redshift. 

\subsection{Merger Rates per halo}

For two  PHBs with masses $m_1$ and $m_2$, the cross section of interaction for binary formation can be expressed as \cite{{Bird:2016dcv}},
\begin{equation}\label{eq:cross-section}
\sigma\left(v_{\textrm{pbh}}\right)=2 \pi\left(\frac{85 \pi}{6 \sqrt{2}}\right)^{2 / 7} \frac{G^2\left(m_1+m_2\right)^{10 / 7} m_1^{2 / 7} m_2^{2 / 7}}{c^{10 / 7} v_{\textrm{pbh}}^{18 / 7}}.
\end{equation}

The differential capture rate is then given by:
\begin{equation}
\frac{d \Gamma_{\textrm{capture}}}{d^3 x}= n_1(r) \cdot n_2(r) \cdot \langle \sigma v_{pbh} \rangle,
\label{eq:diff_capture_rate}
\end{equation}
where $n_1(r)$ and $n_2(r)$ are the number densities of PBHs with mass $m_1$ and $m_2$ respectively. 
We assume that the density distributions for both PBH binaries and single PBHs follow the NFW profile. 
The matter density in PBHs in general is $\rho_{\textrm{PBH}}(r)= f_{\textrm{PBH}} \times \rho_{\textrm{NFW}}(r)$,
with $f_{\textrm{PBH}}$ the fraction of DM in PBHs. In this work, we take $f_{\textrm{PBH}}$ = 1. Our merger rates for the PBH binaries scale as $f_{\textrm{PBH}}^2$.
We note that in this paper we do not aim to revisit the limits on PBHs from the gravitational-wave observations (for a recent update on that see \cite{Franciolini:2022ewd, BAC:2024}), but study the PBH merger rates themselves. We also note that the fraction of PBHs in binaries may depend on $f_{\textrm{PBH}}$ (see \cite{Kavanagh:2018ggo}).

For a monochromatic PBH mass-distribution, i.e. $m_{1} = m_{2} = m$ and consequently we have, 
\begin{equation}
n_{1}\cdot n_{2} \rightarrow \frac{1}{2} n^2 = \frac{1}{2} \left( \frac{\rho_{\textrm{NFW}}(r) \cdot f_{\textrm{PBH}} \cdot f_{m}}{m} \right)^{2}. 
\end{equation}
The factor of 1/2 is used to account for pairs of BHs of identical mass. 
To get the total capture rate per halo, we take integration over the volume of the halo,
\begin{eqnarray}
\Gamma_{\textrm{capture}} &=& 4\pi \int_{0}^{R_{\textrm{vir}}}dr \, r^2 \cdot \langle \sigma \cdot v_{\textrm{pbh}} \rangle  \\ &\times& \frac{1}{2} \frac{ \left[f_{\textrm{PBH}} \cdot \rho_{\textrm{NFW}}(r)\right]^2}{m^2}. \nonumber
\label{eq:Capture_Mono}
\end{eqnarray}
This leads to the final form of the merger rate per halo,
\begin{eqnarray}
\label{Rate_mono}
R_{\textrm{halo}}(M,z)&=&\frac{2 \pi }{3}\left(\frac{85 \pi}{6 \sqrt{2}}\right)^{2 / 7} \cdot f_{\textrm{PBH}}^2 \cdot f_{m}^2 \nonumber \\ 
&\times& \frac{G^2 M_{\textrm{vir}}^2 D(v_{pbh}) f(C)}{R_s^3\cdot  c \cdot  g(C)^2}, 
\end{eqnarray}
where, 
\begin{equation}
f(C)=\left[1-\frac{1}{(1+C)^3}\right]
\end{equation}
and 
\begin{equation}
D(v_{pbh})=\int_0^{v_{\textrm{vir}}} P\left(v_{pbh}, v_{\textrm{disp}}\right)\left(\frac{2 v_{pbh}}{c}\right)^{3 / 7} d v_{pbh}.
\end{equation}

Note that in Eq.(\ref{Rate_mono}), we have substituted for $\rho_{s}$ using Eq.(\ref{rh0_s})
The details for the case of a generic PBH mass-distribution, are provided in Appendix~\ref{App:PBH_mass_distr_2_body}.

\subsection{Direct Capture Rate Results}

\begin{figure}[!htb]
        \centering
        \centering
        \includegraphics[width=\linewidth]
        {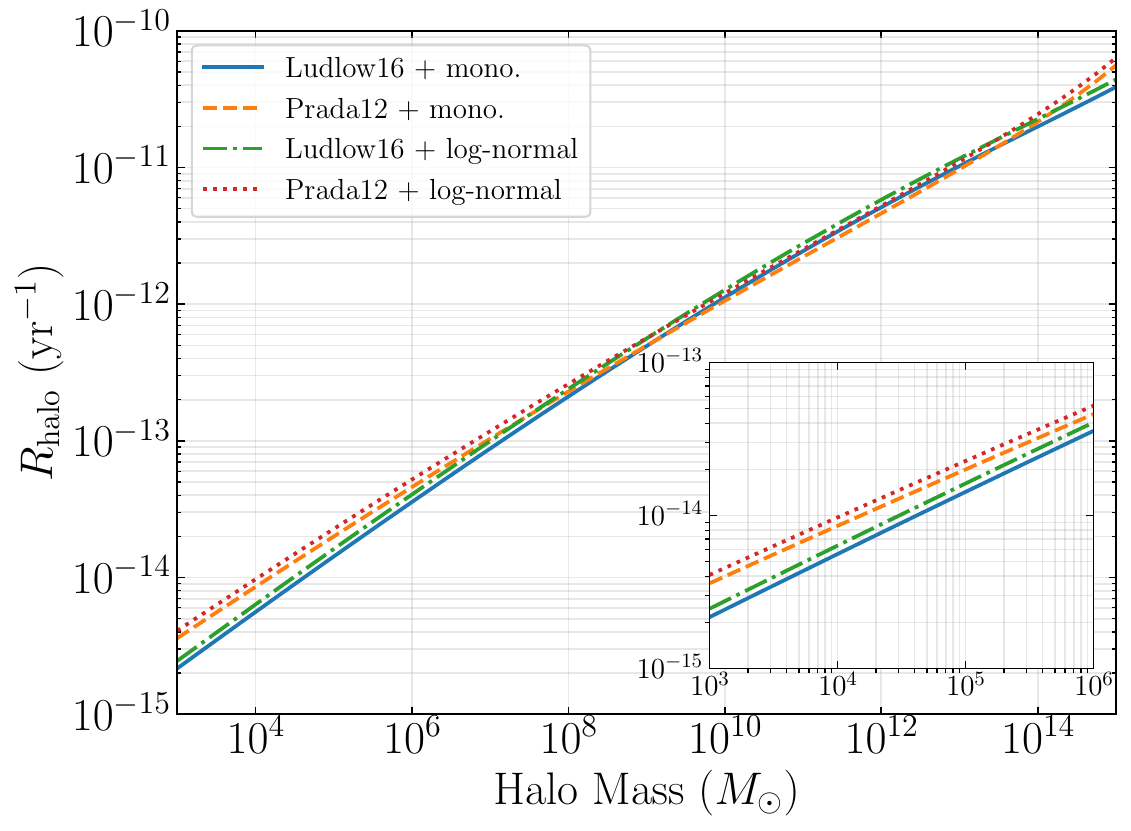}
        \caption{\justifying{The merger rate per halo $R_{\textrm{halo}}$ for the \textit{Ludlow16} and \textit{Prada12} concentration relations. We compare the rates from a monochromatic and a log-normal PBH mass distribution with $\mu = ln (30 M_{\odot})$ 
        and $\sigma =0.6$. We note that in calculating this rates, only 1/2 of the DM halo mass is in single PBHs, while the other 1/2 of the halo mass is in PBH binaries. However, given that most binaries are wide, and act as single black holes with respect to direct captures, our results are independent on the relative abundance of single PBHs vs PBH binaries.}}
        \label{R_halo1}
\end{figure}

In Fig.~\ref{R_halo1}, we present the PBH direct capture rate -and merger rate from that channel- per halo, covering halo masses from $10^3 \, M_\odot$ to $10^{15} \, M_\odot$ at redshift $z=0$. In all our calculations we assume that 50\% of the DM halo mass is in single PBHs and 50\% in PBH binaries. 
We compare the merger rate from direct captures for a monochromatic mass distribution  with $m = 30 \, M_\odot$, to a log-normal mass distribution with mean $\mu = ln(30 M_{\odot})$ 
and a variance $\sigma = 0.6$. 
Changing the value of $m$ for the monochromatic PBH mass distribution does not affect our results as in Eq.~\ref{eq:Capture_Mono} the PBH mass $m$ cancels out. Similarly, changing the value of the mean $\mu$, has a very marginal effect or the capture rates. 
The log-normal mass distribution provides a higher merger rate from two-body captures compared to the monochromatic one.
In Fig.~\ref{R_halo1}, we show results for both the \textit{Ludlow16} and \textit{Prada12} concentration models.
The \textit{Prada12} relation gives higher rates as it predicts higher concentrations than \textit{Ludlow16}. 

\begin{figure}[ht!]
    \centering
    \includegraphics[width=\linewidth]{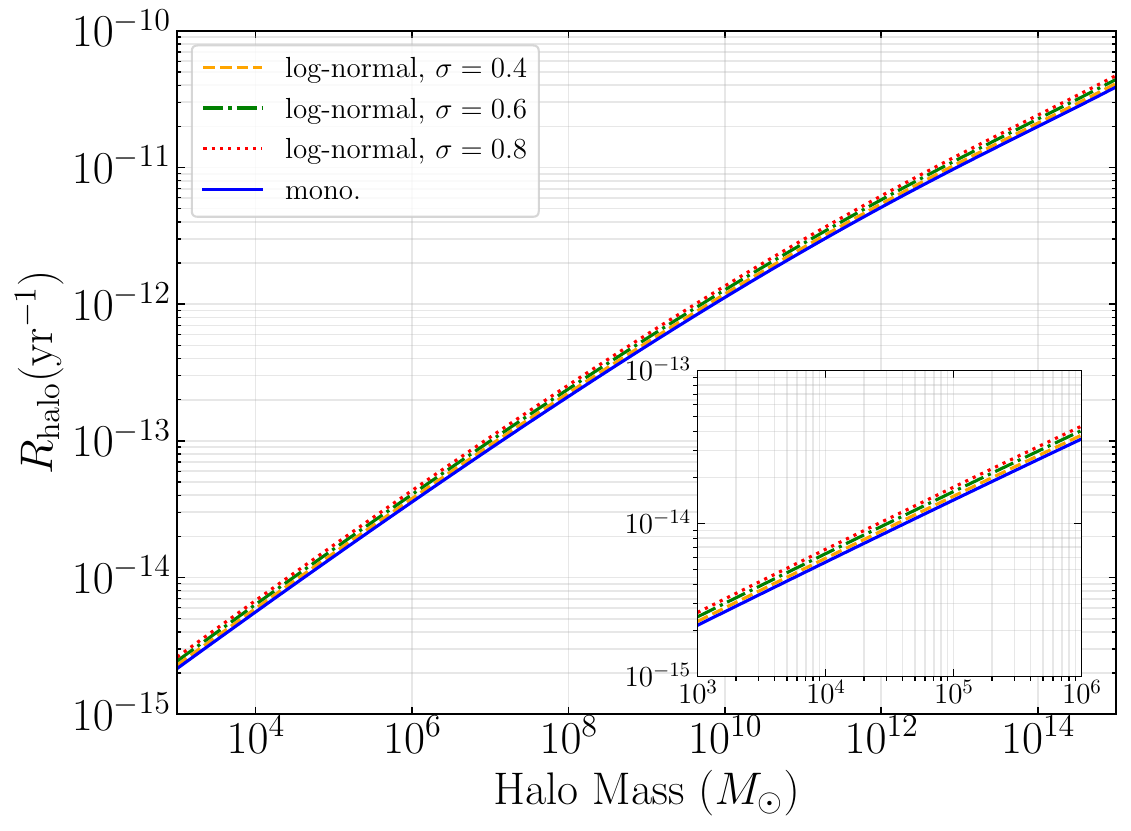}
    \caption{\justifying{The merger rate $R_{\textrm{halo}}$ per halo at redshift $z=0$, considering the \textit{Ludlow16} concentration relation. We show results for both a monochromatic and a log-normal PBH mass distribution with $\mu= ln (30 M_{\odot})$. }}
    \label{R_halo2}
\end{figure}

In Fig.~\ref{R_halo2}, we show the merger rate $R_{\textrm{halo}}$ as a function of halo mass at redshift $z=0$ for the \textit{Ludlow16} concentration relation. 
The plot shows both the monochromatic and log-normal distributions with mean $\mu= ln (30 M_{\odot})$ and $\sigma=[0.4, 0.6, 0.8]$. 
As we increase $\sigma$ in the log-normal distribution, the merger rate becomes higher.

\begin{figure}[ht!]
    \centering
    \includegraphics[width= \linewidth]{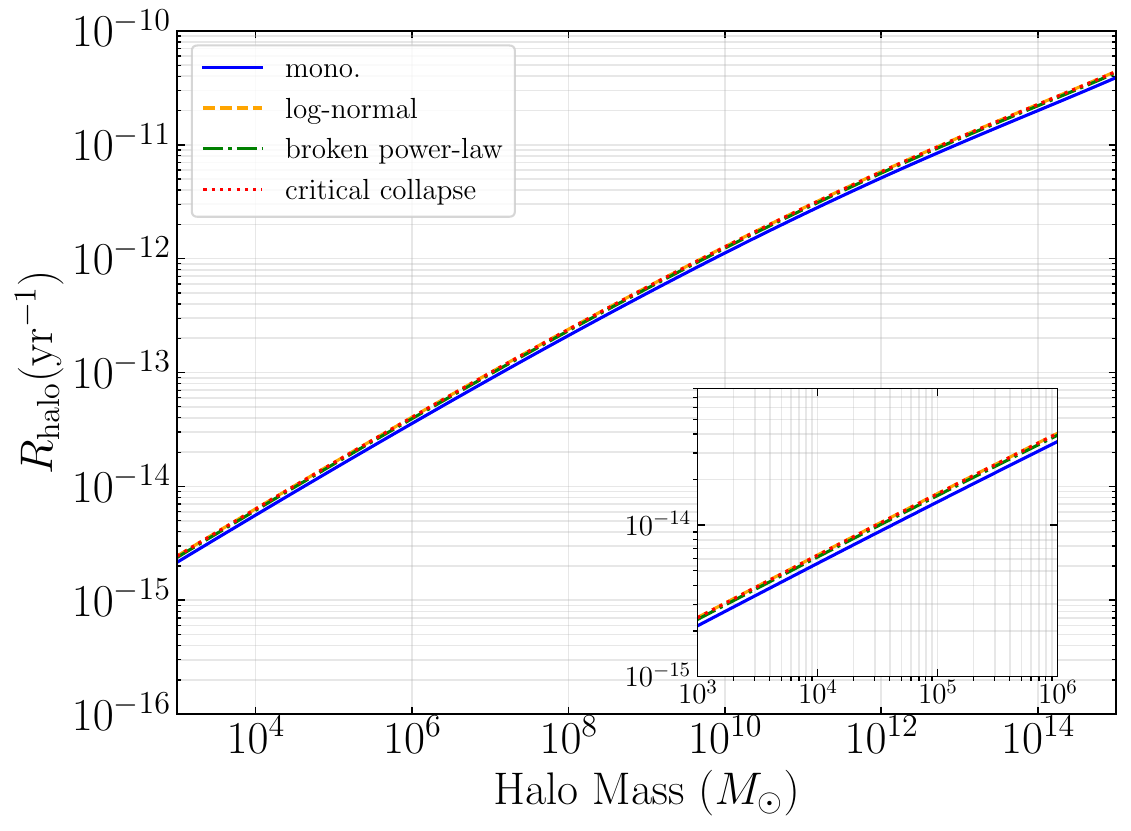}
    \caption{\justifying {The merger rate per halo $R_{\textrm{halo}}$ per halo at redshift $z=0$ considering \textit{Ludlow16} concentrations model consdiering different mass distributions of PBHs, including monochromatic, log-normal, critical collapse and broken power law mass functions.}}
    \label{R_halo3}
\end{figure}

 In Fig.~\ref{R_halo3}, we compare merger rates using monochromatic mass distributions with log-normal and critical collapse distributions. Critical collapse models predict slightly higher merger rates than the log-normal or monochromatic mass-distributions. The three PBH mass distributions are shown in the top panel of Fig.~\ref{fig:PBH_PDF} in Appendix~\ref{PBH_PDFs}. 

\begin{figure}[ht!]
    \centering
    \includegraphics[width=\linewidth]{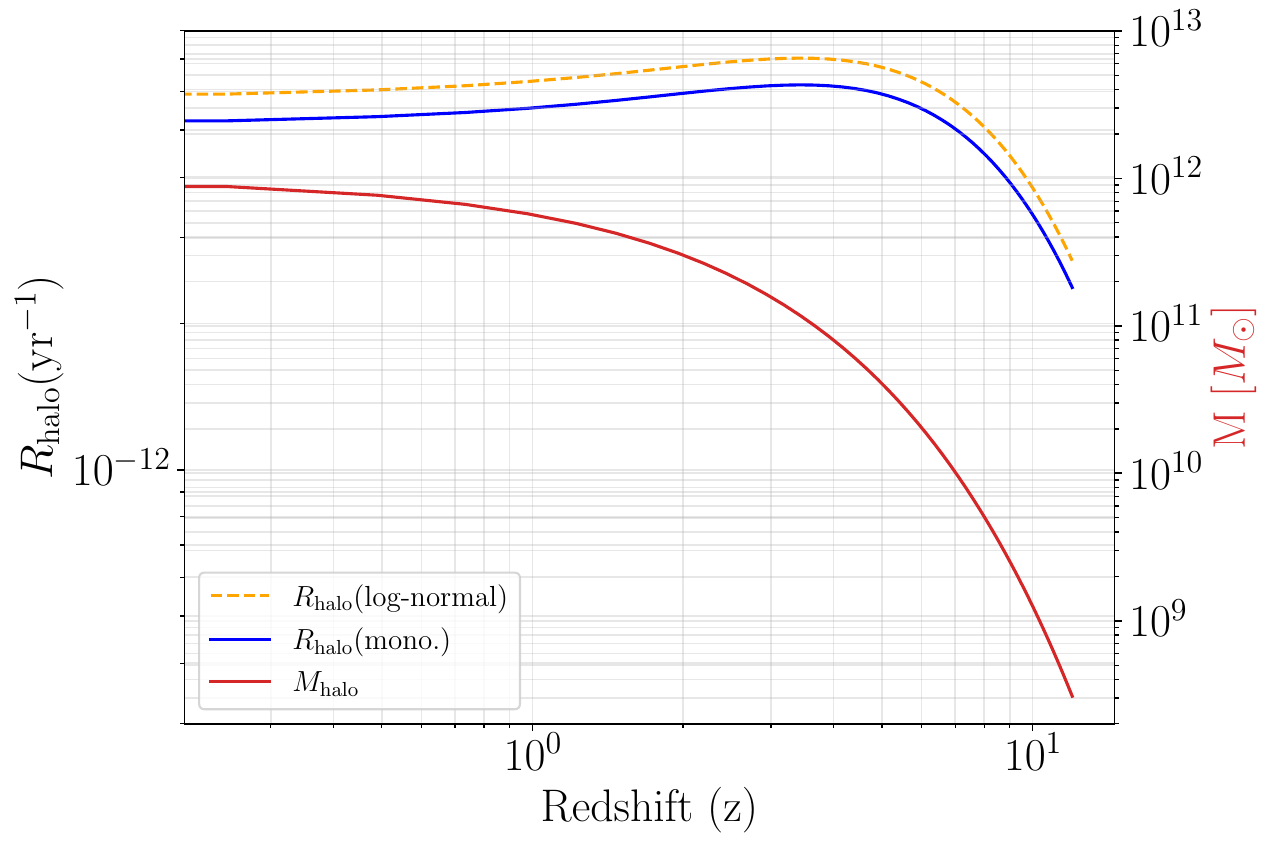}
  \caption{\justifying  Evolution of the merger rate per halo $R_{\textrm{halo}}(z)$  and halo mass with redshift for a halo having a mass $M=10^{12}M_\odot$ at $z=0$. The \textit{Ludlow16} concentration model is applied along with log-normal distribution (with $\mu=ln(30 \, M\odot)$ and $\sigma=0.6$) for the PBH mass functions.}  
    \label{fig:R_halo_evol}
\end{figure}

In Fig.~\ref{fig:R_halo_evol}, we show the redshift evolution of the merger rate due to captures of two PBHs inside a halo with a mass that at $z=0$ is $10^{12}M_\odot$. For comparison we also show how such a halo's mass evolved (same red line as in Fig.~\ref{fig:CONC_MASS_Z}).
We utilize the \textit{Ludlow16} concentration model and  show results for both the monochromatic PBH mass distribution and the log-normal distribution. 
For such a DM halo, the PBH merger rate from direct capture events increased with time peaked at $z\simeq5$, after which time it decreased asymptotically to a constant rate. 
The log-normal distribution provides higher merger rates than the monochromatic by a factor of 3 for $z<6$.
For the evolution of the halo mass, we used Eq.(\ref{eq:MAH}) 

\begin{figure}[ht!]
    \centering
    \includegraphics[width=\linewidth]
{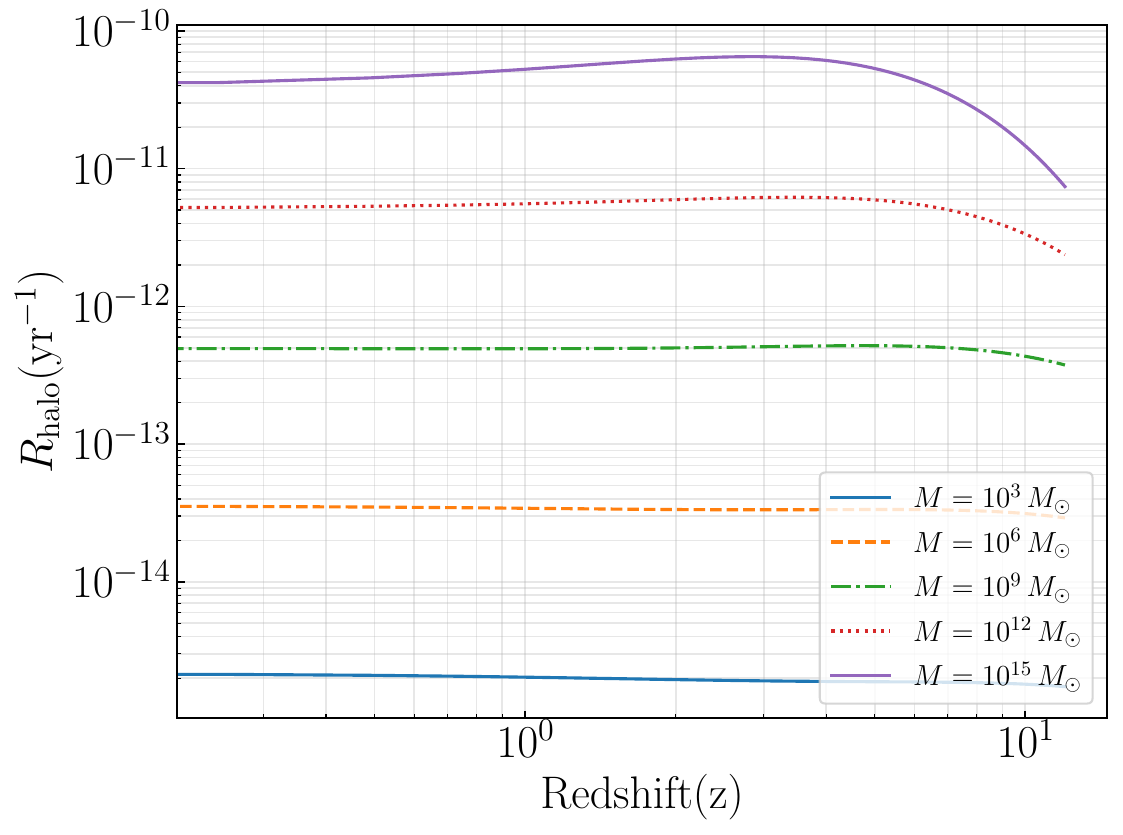}
    \caption{\justifying Evolution of the merger rate per halo, $R_{\textrm{halo}}(z)$, with redshift $z$ for halos of masses $10^3$, $10^6$, $10^9$, $10^{12}$, and $10^{15} M_\odot$ at $z=0$. The \textit{Ludlow16} concentration model has been used, incorporating the monochromatic PBH mass with $m_{PBH}=30$.}
    \label{R_halo_evol_masses}
\end{figure}

Similarly, in Fig.~\ref{R_halo_evol_masses}, we compare the redshift evolution of the capture merger rate per halo $R_{\textrm{halo}}(z)$ for various halos, specifically $10^3$, $10^6$, $10^9$, $10^{12}$, and $10^{15} M_\odot$ halos at $z=0$. 
For smaller mass halos, the merger rate from direct capture events peaks at later times.

\begin{figure}[ht!]
    \centering
    \includegraphics[width=\linewidth]{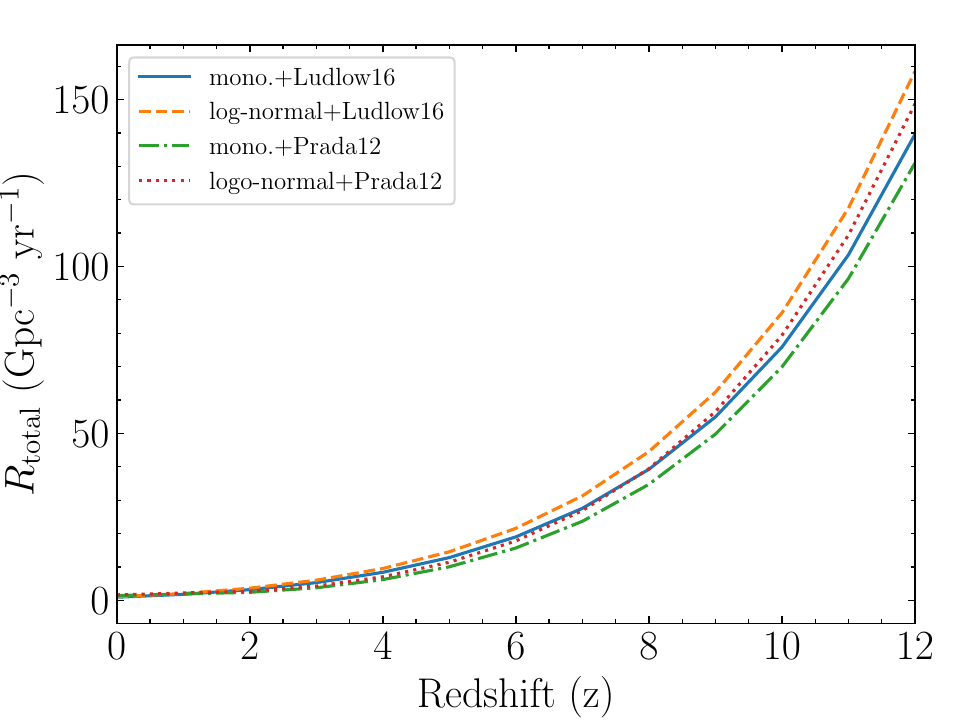} 
    \caption{\justifying {The redshift evolution of the comoving PBH merger rate from two-body captures. We used the \textit{Press-Schechter} halo mass function (see text for details). For the log-normal PBH mass distribution, we have used $\mu= ln( 30 \, M_\odot)$ and variance of $\sigma=0.6$.}}
    \label{R2b_total_PS}
\end{figure}

Fig.~\ref{R2b_total_PS}, shows the total comoving merger rate with the \textit{Press-Schechter} halo mass function versus redshift. This rate is evaluated by combining the merger rate per halo with halo mass function,
\begin{equation}
R(M,z)=\int R_{{halo }}(M,z) \frac{d n}{d  M} d M .
\end{equation}
We show results for two PBH mass distributions: monochromatic and log-normal. For the computation of the halo mass function we use \texttt{HMFcalc}. Additionally, we employed both the \textit{Ludlow16} and \textit{Prada12} concentration models in our calculations. We illustrate the differences in merger rates predicted by these models and mass functions. We observe good agreement between the alternative assumptions for $z\lesssim 2$. 
However, at higher redshifts, uncertainty arises due to the mass function of the halos. The \textit{Prada12} concentration model predicts higher concentrations than \textit{Ludlow16} at low redshifts. This leads to a slightly higher total merger rate using the \textit{Prada12} to the \textit{Ludlow16} for $z \lesssim 2$.
However, in Fig.~\ref{R2b_total_PS}, at $z > 2$ the \textit{Ludlow16} relation predicts higher merger rates. 
This can be attributed to the behavior of the 
\textit{Prada12} relationship for massive halos at $z>2$. 
For these halos and redshifts the \textit{Prada12} relationship diverges, giving in fact increasing concentrations with increasing redshift. 
As a result going back in time we had to 
cup the predicted concentrations to from the \textit{Prada12} model to their minimum values. We consider that to simply be part of the generic modeling uncertainty for this calculation. 

\section{Binary- single interactions events}\label{three_body}
In this section, we outline our assumptions to calculate the merger rate resulting from binary-single interactions. 
There are two types of interactions that PBH binaries encounter in DM halos : a) interactions between PBH binaries and single PBHs, and b) interactions between PBH binaries themselves. 

In this study, we care about the PBH binaries that are hard at the beginning of the evolution of their orbital properties and can merge within a Hubble time. See appendix \ref{app:Prestime_PBHbinaries} to  see  what fraction of the binaries that are hard per halo at $z=0$. These are the binaries whose semi-major axis is less or equal to the critical value of the semi-major axis in a halo. We also take for simplicity a monochromatic mass distribution with $m=30 M_{\odot}$.
Due to the wide distribution in the orbital properties of the PBH binaries, the interactions between two PBH binaries can be approximated well as the interaction between the harder of the two binaries and the PBH closest to it. 
Thus, we treat all interactions of hard PBH binaries with their environment, as interactions with single PBHs. Because all PBHs are taken to have the same mass, there are no exchange interactions. For narrow PBH mass-distributions this is a valid approximation.

\subsection{The evolution of PBH binaries properties}
To model the merger rate of PBH binaries through binary-single interactions, we examine their orbital parameters $(a,e)$ evolution equations. 
For a PBH binary with masses $m_1$ and $m_2$ residing in a DM halo, the evolution equation for its semi-major axis is \cite{Peters},

\begin{eqnarray}
\label{eq:peters}
 \frac{da}{dt}&=&-\frac{G H \rho_{\textrm{env}}(r,t)}{v_{\textrm{disp}}^{\textrm{env}}(r,t)} a^2-\frac{64}{5} \frac{G^3}{c^5 a^3} \nonumber \\
&&\times \left(m_1+m_2\right) \cdot\left(m_1 \cdot m_2\right)F(e),
\end{eqnarray}
with, 
\begin{equation}
\label{eq:Fe_func}
F(e)=\left(1-e^2\right)^{-7 / 2} \cdot\left(1+\frac{73}{24} e^2+\frac{37}{96} e^4\right).
\end{equation}
$G$ is Newton's constant, $c$ is the speed of light, and $\rho_{\textrm{env}}(r,t)$ is the density of the environment in which the PBH binary resides, a DM halo in our case. 
Finally, $v_{\textrm{disp}}^{\textrm{env}}(r,t)$, is the velocity dispersion of the PBHs surrounding the binary.

For any halo under consideration, we assume that the density distributions for both PBH binaries and single PBHs adhere to the NFW profile. This implies that the environmental density in Eqs.(\ref{eq:peters},~\ref{eq:Fe_func}) is determined according to the NFW profile.
We assume that $50\%$ of the halo's mass is in PBH binaries and the remaining $50\%$ in single PBHs. 
Consequently, 
\begin{eqnarray}\label{assumption}
\rho_{\textrm{env}}(r,t) &=&  \rho_{\textrm{PBH \, binaries}}(r,t) + \rho_{\textrm{single \, PBH}}(r,t) \nonumber \\
&=& \rho_{\textrm{NFW}}(r,t).
\end{eqnarray}
We remind the reader that the first term in Eq.(\ref{eq:peters}) describes the averaged effect of hardening interactions of PBH binaries with their environment, while the second term represents the Peters secular evolution due to gravitational wave (GW) emission. 
In Eq.~\ref{assumption} for the $\rho_{\textrm{env}}$ we include the density of the binaries $\rho_{\textrm{PBH \, binaries}}$, as the great majority of PBH binaries are wide and act as two single objects on the evolution of the tight binaries.
$H(r,t)$ is the hardening rate~\cite{H_rate}, which can be approximated by \cite{H_approx},
\begin{equation}
H(r,t)=14.55 \times\left(1+0.287 \frac{a}{a_h(r,t)}\right)^{-0.95}.
\end{equation}
The hard semi-major axis $a_h(r,t)$ is defined by,
\begin{equation}
a_h(r,t)=\frac{G m_{1}}{4 v_{\textrm{disp}}^{\textrm{env}}(r,t)^{2}}.
\end{equation}
The velocity dispersion of single objects (PBHs) and binaries in the halo depends on time and position,
\begin{equation}
v_{\textrm{disp}}^{\textrm{env}}(r,t)
(r,t)=\sqrt{\frac{2 G M(r,t)}{r(t)}},
\end{equation}
where $M(r,t)$ is the halo mass contained within a sphere of radius $r$ from its center, at time $t$. 

The evolution equation for eccentricity of the PBH binaries is \cite{Peters},
\begin{eqnarray}
\label{eq:ecc}
 \frac{de}{dt}&=&\frac{G H(r,t) K(r,t) \rho_{\textrm{env}}(r,t)}{v_{\textrm{disp}}^{\textrm{env}}(r,t)} a-\frac{304}{15} \frac{G^3}{c^5 a^4}     \nonumber \\
&&\times \left(m_1+m_2\right)\left(m_1 \cdot m_2\right) D(e),
\end{eqnarray}
with,
\begin{equation}
D(e)=\left(1-e^2\right)^{-5 / 2} \cdot\left(e+\frac{121}{304} e^3\right).
\end{equation}
$K(r,t)$ describes the eccentricity growth rate as a result of interactions with third bodies. Like $H(r,t)$, it is determined by numerical three-body simulations. We use the fitting function provided in \cite{H_approx} (their Eq. 18). 

Our simulations neglect the possibility for PBH ejection from DM halos during binary-single interactions. Given that the escape velocities from massive halos are generally high, except in the outermost regions, these ejections have a minimal impact on the merger rates. However, in the case of small mass halos, such ejections may reduce the merger rate within the halo, as the escape velocity is comparable to the dispersion velocity. Consequently, neglecting these ejections could lead to an overestimation of merger rates in small mass halos.
To properly account for the significance of ejections in small mass DM halos, numerical simulations are required. This goes beyond the scope of this work. 

\subsection{Numerical Analysis Setup}
\label{subsec:numerical_setup_binary_single}
The system Eqs.(\ref{eq:peters},~\ref{eq:ecc}) is solved numerically. 
We divide our halos with mass larger than $10^{7} \, M_{\odot}$ into ten discrete spherical shells and take the PBH binaries to reside inside 
a given shell throughout the simulation.
The radial boundaries of these shells $R_i$ are logarithmically spaced from $0$ to $R_{\textrm{vir}}(t)$.
This allows for the shells to grow with time following the mass and volume evolution of the DM halos.
For a halo with $R_{\textrm{vir}}(t)$ at a given time $t$, the radial boundary $R_{i}(t)$ is,
\begin{equation}
\label{R_I}
R_{i}(t) = \exp\left( \frac{i}{N_{\textrm{shell}}} \cdot \ln \left(1 + \frac{R_{\textrm{vir}}(t)}{1 \, \textrm{pc}}\right) \right) - 1,
\end{equation}
where $i \in [0,10]$.

For less massive halos, we use the same approach as described previously, with the difference of using a smaller number of shells. For halos of $O(10^{3}) M_{\odot}$ we only take one sphere. For $O(10^{4}) M_{\odot}$ one sphere and one shell around it. For $O(10^{5}) M_{\odot}$ one sphere and two logarithmically spaced shells around it ($i \in [0,3]$) and for $O(10^{6}) M_{\odot}$ $i \in [0,5]$. As an example, for a halo with a mass of $10^{12} M_\odot$ and virial radius of $R_{\textrm{vir}}$$\simeq211$ kpc at $z=0$, the shell boundaries at the end of the simulation are given in Table~\ref{Ri_values}.
\begin{table}[ht!]
\centering
\begin{tabular}{|c|c|}
\hline
$ i $ & $ R_i (\textrm{pc}) $ \\
\hline
0 & 0.00 \\
1 & 2.41 \\
2 & 10.61 \\
3 & 38.57 \\
4 & 133.83 \\
5 & 458.46 \\
6 & 1564.65 \\
7 & 5334.13 \\
8 & 18179.03 \\
9 & 61949.37 \\
10 & 211101.44 \\
\hline
\end{tabular}
\caption{\justifying{The shell boundaries $R_{i}$ at the end of the evolution of a DM halo with $M=10^{12} M_\odot$ and $R_{\textrm{vir}}=\simeq211$ kpc at $z=0$.}}
\label{Ri_values}
\end{table}

Furthermore, we take the halo density $\rho_{\textrm{env}}(r,t)$ and the dispersion velocity $v_{\textrm{disp}}^{\textrm{env}}(r,t)$ to evolve in discrete timesteps. We update their values every $200$ Myr. 

The densities on each spherical shell are evaluated at the midpoint of the shell. 
Thus, the density of the shell is,
\begin{equation}
\rho_{\textrm{i}}(t) = \rho_{\textrm{NFW}}\left(\frac{R_{i}(t) + R_{i+1}(t)}{2}\right).
\end{equation}

We remind the reader that $R_{\textrm{vir}}=R_{s} \cdot C(M,z)$ and that $R_{\textrm{s}}$ depends on $C(M,z)$. 
We employ the \textit{Ludlow16} concentration relation and evaluate the halo's mass at any given redshift. 
Once we have the mass at a specific redshift, we can estimate $C(M,z)$ and also derive parameters such as $R_{\textrm{vir}}(t)$.

The system of Eqs. (\ref{eq:peters},~\ref{eq:ecc}) is solved using the Euler method for each PBH
binary,
located within a given shell at each timestep in the history of DM halo. 
We implement structured time-stepping to ensure precision and efficiency. We adapt a global timestep $ dt_{\textrm{global}} = 200 $ Myr, that governs the main simulation time $t$ from 0 to $t_{\textrm{max}}$, the look-back time from when we start our simulation to the present day 
\footnote{For DM halos that at the present era have a mass 
$M \gtrsim 5 \times 10^{4} \, M_{\odot}$, we start our simulation at $z=12$. Instead, for halos with smaller masses at the present era, we start their evolution at a lower redshift, as we require that halos at all times contain at least $\simeq 30$ PBHs in them.}.  Within each $dt_{\textrm{global}} = 200$ Myr we evolve numerically the
binaries' orbital properties using a timestep of $ dt_{\textrm{local}} =2$ Myr.  
During each global timestep $ dt_{\textrm{global}}$ , we evolve our hard PBH binaries through:
\begin{eqnarray}
    a_{n+1} &= a_n + dt_{\textrm{local}} \cdot f_a(t_{\textrm{local}, n}, a_n, e_n), \\
    e_{n+1} &= e_n + dt_{\textrm{local}} \cdot f_e(t_{\textrm{local}, n}, a_n, e_n),
\end{eqnarray}
where $ a_n $ denotes the semi-major axis and $ e_n $ represents the orbital eccentricity at step $n$. The functions $ f_a $ and $ f_e $ represent the right hand side of Eqs.~(\ref{eq:peters},~\ref{eq:ecc}). 
If during the evolution of the semi-major axis of our PBH binaries $a$ approaches zero, we consider that a merger event. 

\subsection{Evolution of the DM halo's mass properties}

To evolve the mass of DM halos, we used the semi-analytical model of Ref.~\cite{MAH} (their Appendix C), which predicts that,
\begin{equation}
 M(z) = M_0 (1+z)^\alpha e^{\beta z}. 
 \label{eq:MAH}
\end{equation}
Parameters $\alpha$ and $\beta$ characterize the mass evolution of each halo. 
Eq.~(\ref{eq:MAH}) is used to determine the concentration parameter $C(M, z)$ of a given halo at a given redshift in its evolution and subsequently the $R_{\textrm{vir}}(t)$. The details on how to calculate these parameters are given in appendix(\ref{MAH_params}). 

\begin{figure}[ht!]
    \centering
    \includegraphics[width=\linewidth]{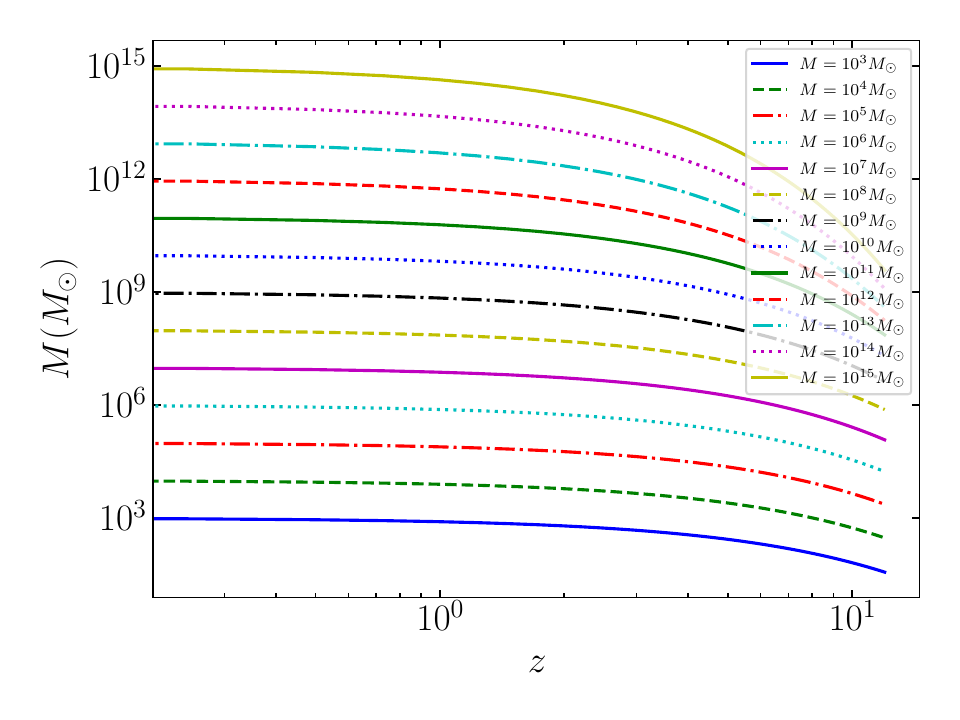}
    \caption{\justifying Mass accretion history for a range of halos mass with  $10^{3} -10^{15} \, M_\odot$ at present day, $z=0$.}
    \label{fig:MAH1}
\end{figure}

In Fig.~\ref{fig:MAH1}, using Eq.~\ref{eq:MAH} 
we show the mass evolution starting from $z=12$, of halos with masses of $10^{3} -10^{15} M_\odot$ at $z=0$. We observe the same trend as for the case of a $10^{12}  M_\odot$ halo (see Fig.~\ref{fig:CONC_MASS_Z}). Low-mass halos form earlier but gain their mass more slowly compared to massive halos.

\begin{figure}[ht!]
    \centering
    \begin{subfigure}[b]{\linewidth}
        \centering
        \includegraphics[width=\linewidth]{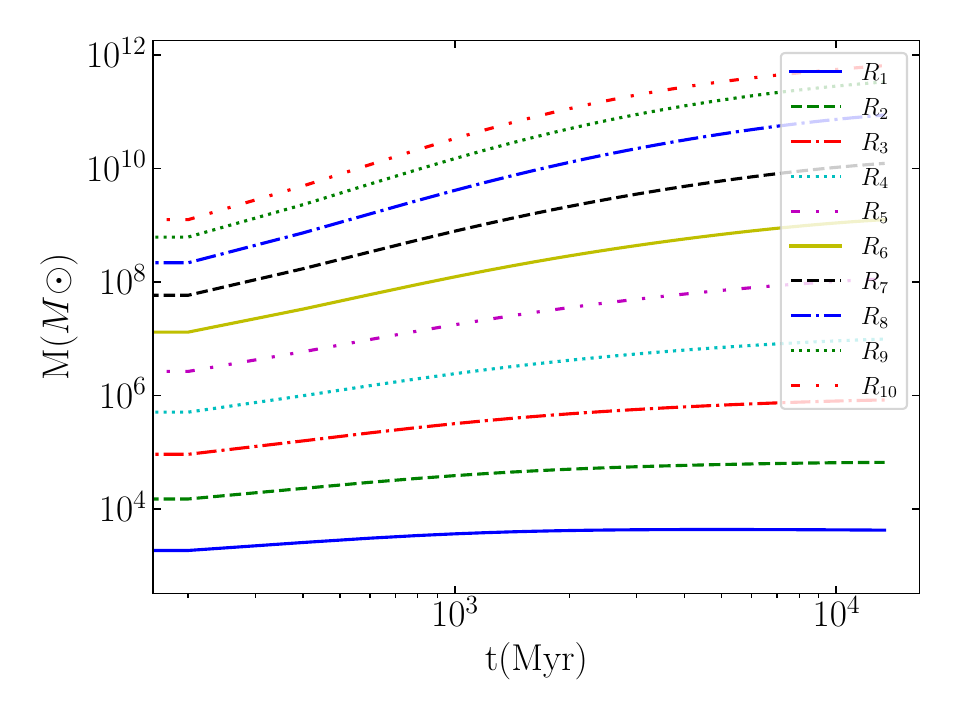}
        \label{}
    \end{subfigure}
    
    \begin{subfigure}[b]{\linewidth}
        \centering
        \includegraphics[width=\linewidth]{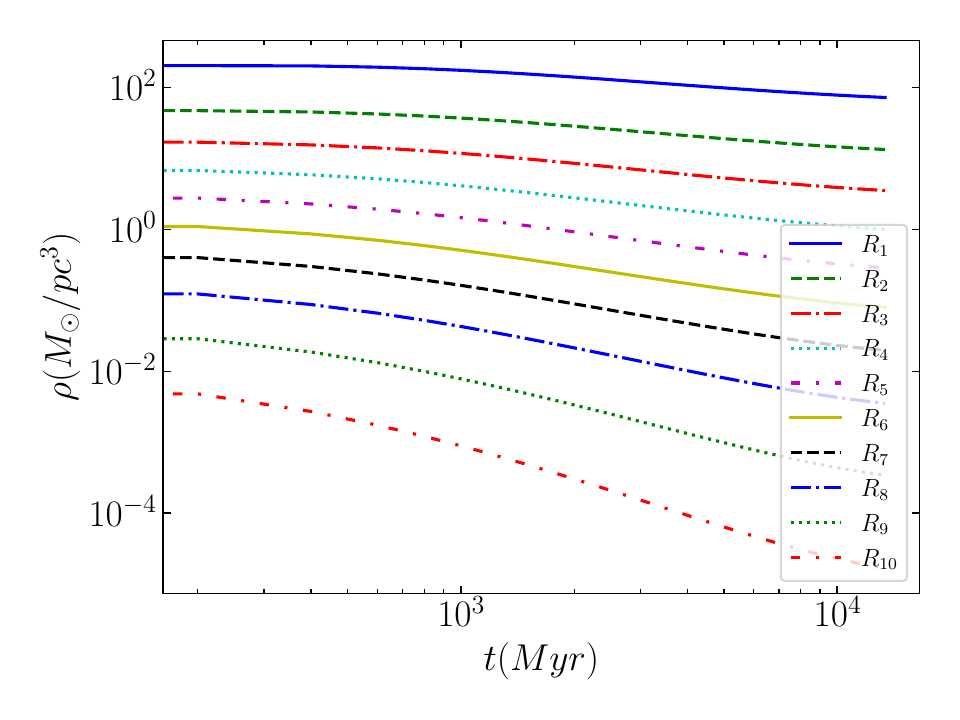}
        \label{}
    \end{subfigure}
    
    \caption{\justifying Evolution of mass (\textit{top}) and density of a halo (\textit{bottom}) with $M=10^{12} M_\odot$ in each spherical shell with radius $R_{i}(t)$ where $i=1,2,3,\ldots,10$.}
    \label{fig:halo_evolution}
\end{figure}

In Fig.~\ref{fig:halo_evolution}, we depict the time evolution of the mass and density for each shell of a halo with a present-era mass of $10^{12} M_{\odot}$. We observe that the density is higher in the inner shell compared to the outer ones. However, due to the relevant volumes of the spherical shells, the mass of the outer shells is much larger. 

\begin{figure}[ht!]
    \centering
    \begin{minipage}[b]{0.99\linewidth}
        \centering
        \includegraphics[width=\linewidth]{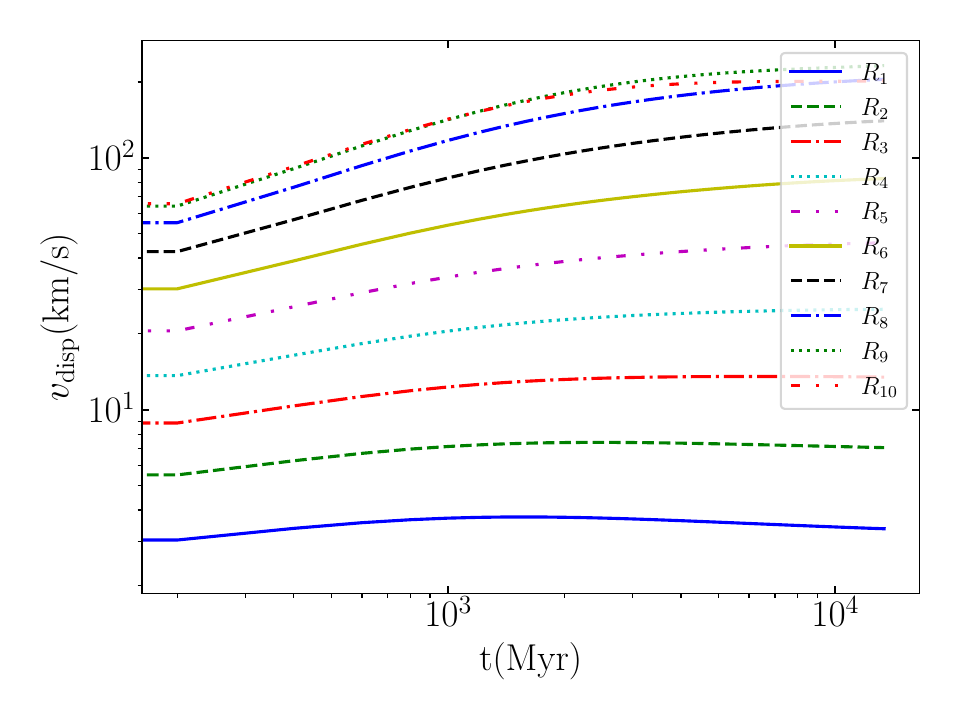}
        \label{}
    \end{minipage}\hfill
    \begin{minipage}[b]{0.99\linewidth}
        \centering
        \includegraphics[width=\linewidth]{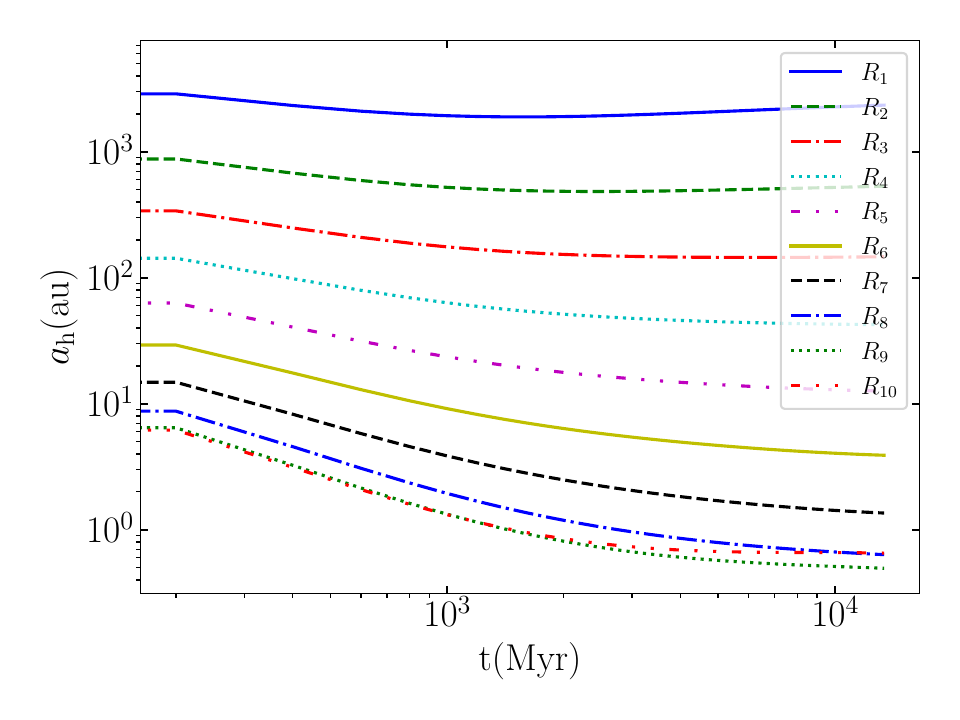}
        \label{}
    \end{minipage}
    
    \caption{\justifying The time evolution of  the velocity dispersion $v_{\textrm{disp}}$ (top panel) and $a_{h}$ (bottom panel) for each of the ten spherical shells of a halo with $M = 10^{12} M_\odot$ at the present era.}
    \label{v_disp_a_hard}
\end{figure}

Fig.~\ref{v_disp_a_hard} top panel, shows the evolution of dispersion velocity $v_{\textrm{disp}}(r,t)$  in each spherical shell of a $10^{12} \, M_{\odot}$ halo. 
Furthermore, the bottom panel shows the critical value for the semi-major axis of hard binaries $a_{h}$, for the different spherical shells. We observe an opposite trend to the velocity dispersion as $a_{h}=\frac{G m}{4 v_{\textrm{disp}}^2}$ ($m$ being the PBH mass). 

We note that we only evolve the hard PBH binaries. Each time, we sample a specific number of PBH binaries and then select the hard binaries using the condition $a \leq a_{h}$. 
Only hard binaries may merge as a result of interactions with third bodies \cite{Heggie}. 
From the bottom panel of Fig.~\ref{v_disp_a_hard}, we notice that the inner mass shells have relatively constant critical values for $a_{h}$. However, in the outer shells $a_{h}$ decreases with time. 
For the halo we simulate here, i.e. $M=10^{12} \, M_{\odot}$ at $z=0$, at early times there was a significantly higher fraction of hard binaries than later. 
This has implications on the resulting merger rate evolution, as will be shown in the following sections. 

\subsection{The Initial distribution of PBH binaries' orbital properties}

PBH binaries exist even outside DM halos.
We assume that these binaries are gravitationally bound and remain unaffected until the formation of DM halos. 
Therefore, at the time they become part of a DM halo, their initial orbital parameters $(\alpha_{0}, e_{0})$ follow the same distributions as at the time for their formation at matter-radiation equality. 
Only, once these binaries fall into DM halos, their evolution is affected by their environment.
In each time-step of our simulation, we pick $N_{\textrm{sample}}$ of PBH binaries with ($a_0$, $e_0$) to evolve between $t$  and $t+dt$. 
To sample these parameters, we take the approach presented in \cite{PBBH1,PBBH2,PBBH3}. 
In Appendix~\ref{app:Prestime_PBHbinaries}, we provide more details on how we sample the initial orbital parameters of the PBH binaries.

\subsection{Binary-single interaction results}

In this section, we present the results of our simulations for the merger rates per halo (in $\textrm{yr}^{-1}$) and the total merger rate (in $\textrm{Gpc}^{-3}\cdot \textrm{yr}^{-1}$) from the interactions of PBH binaries with single PBHs. There are also interactions of the PBH binaries with regualr starts. Those too can cause hardening of the PBH binaries' orbital properties. However, these happen at the cores of the most massive halos, which as we will show contribute very little to the total rate. Thus, we can ignore the impact of stars.

In our simulations for a given mass halo, we have assumed a monochromatic mass distribution $m=30 \, M_{\odot}$ and used the \textit{Ludlow16} concentration model, under the assumption of Eq.(\ref{assumption}) that $50\%$ of the halo's mass is in PBH binaries and the remaining $50\%$ in single PBHs. For a given halo, we start the simulation at $z=12$ and end around $z=0$ which s equivalent simulation time of 0 to 13.4 Gyr.

\begin{figure}[ht!]
    \centering
    \includegraphics[width=\linewidth]{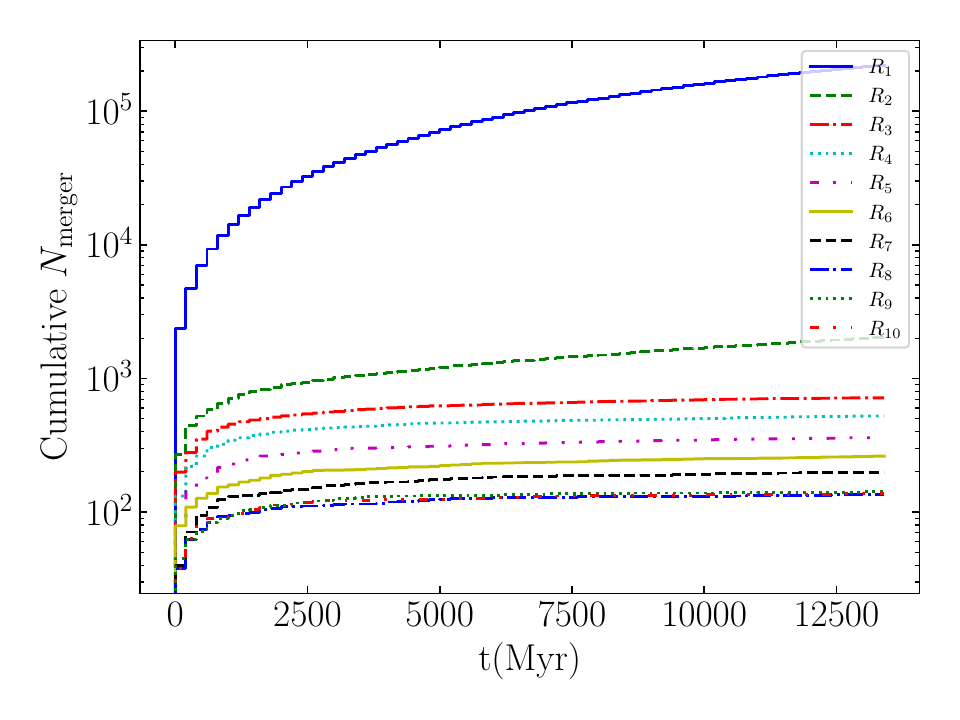}
\caption{\justifying The cumulative number of PBH mergers from binary-single interactions in each mass shell $R_{i}$ of a $1.15 \times 10^{12} \, M_{\odot}$ halo. In the simulation, in each shell we have sampled $N_{\textrm{sample}}=2\times 10^{6}$  binaries per  time-step.}
    \label{N_merger_cum_v1}
\end{figure}

In Fig.~\ref{N_merger_cum_v1}, we present the cumulative number of binary PBH mergers $N_{\textrm{merger}}$ in each of the shells $R_{i}$ of a halo $1.15 \times 10^{12} M_\odot$ over time. 
As a first step in our simulation, we take that there are 
$N_{\textrm{sample}}=2\times 10^{6}$  binaries in each shell and in each time-step. That is obviously not accurate as there are many more PBH binaries in the outer shells and at later times. We describe how we re-scale our results subsequently.  
Before re-scaling, we see that the cumulative number of mergers is higher in the inner shells. In the inner shells, the PBH density is much higher while their velocity dispersion lower. Thus, the PBH binaries experience hardening more often, evolving much faster and giving a much larger number of mergers. In Appendix~\ref{Timescale_app}, we present information on how fast hard PBH binaries are expected to have hardening interactions with single PBHs for different DM mass halos and at different times in those halo's evolution. We find that for the smaller in mass halos and at later times in their evolution the interactions of PBH binaries with single PBHs are important.  

The true merger rate of PBH binaries due to their interaction with single PBHs, is derived by taking the results of Fig.~\ref{N_merger_cum_v1}, and re-scaling them by accounting for the true number of binaries that exist per time-step, using,
\begin{equation}
R_{\textrm{halo}} = \sum_{i=1}^{N_{\textrm{shells}}} \sum_{t=0}^{t_{\textrm{look}}} \frac{N_{\textrm{PBH  \, binaries}, i, t}}{N_{\textrm{sample}}} \cdot 
\frac{N_{\text{merger}, i, t}}{\Delta t}
.
\end{equation}
$t_{look}$ is the look-back time of a halo, $N_{\textrm{BBH}, i, t}$ is actual number of binaries in a shell with radius $R_{i}$ from halo's center at $t$, $N_{\textrm{merger},i,t}$ is the number of mergers in the shell at time $t$ when $N_{\textrm{sample}}$ binaries are considered in the simulations, and $N_{\textrm{sample}}$ is the number of binaries sampled per time-step. 

After re-scaling, the merger rates in the inner shells are significantly suppressed due to much fewer binaries in them compared to the outer shells. We also find that most mergers occur early in the evolution of the halo. At high redshifts, the density of single PBHs and PBH binaries are high and the velocity dispersion low, leading to a higher hardening rates and consequently more mergers in early times.

\begin{figure}[ht!]
    \centering
        \includegraphics[width=\linewidth]{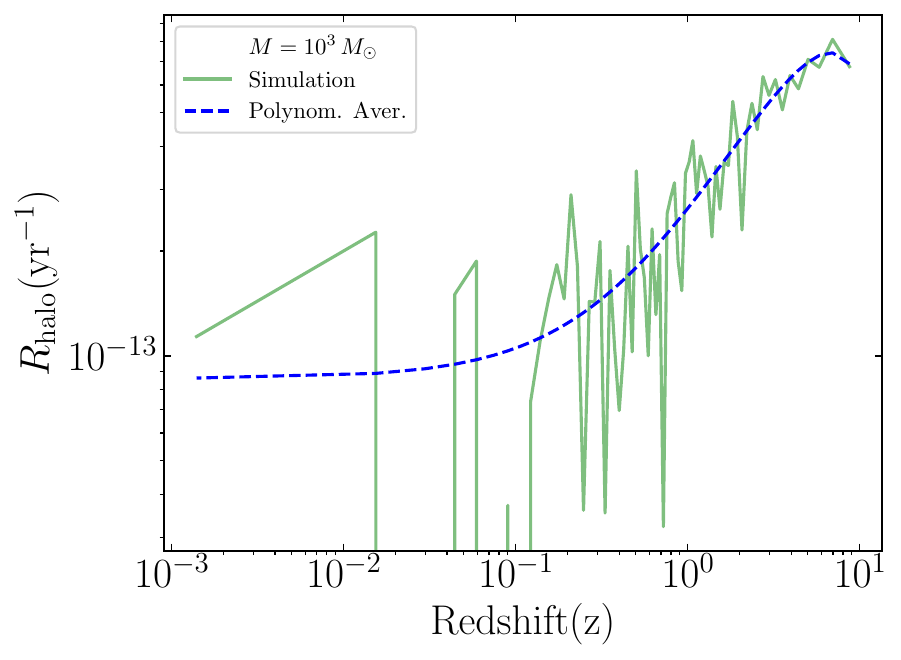}
        \caption{\justifying The merger rate per halo from binary-single interactions for a halo with mass $M=10^{3} \, M_{\odot}$ at z=0. The green spiky line shows the results from our simulation, the dashed blue (``Polynom. Aver.'') line is a polynomial averaging to the simulation. Our simulations start at redshift at $z=12$ and stop at $z=0$ corresponding to the lock back time of $t=13.4$ Gyr. However, we drop the first time-step from the calculation of the rates. So our results start at $z\simeq 10$. The simulation has been done assuming the halo density and dispersion velocity is calculated at  $R_{\textrm{vir}}/2$ at each time-step. }
    \label{3body_rate_1e3}
\end{figure}

In Fig.~\ref{3body_rate_1e3}, we present the merger rate per halo for $10^3 \, M_{\odot}$, halo where the solid green spiky line depicts the merger rate obtained from the simulation that has stochastic fluctuations and the dashed blue line represents a polynomial approximation (averaging) of it. 
While our simulation starts at $z=12$, since we are averaging rates between neighboring time-steps, we drop the first time-step from the averaged calculation of the rate and present results starting from $z\simeq 10$.
Especially for the case of $O(10^{3}) \, M_{\odot}$ DM halos, there is no velocity gradient for the PBHs, as the time-scale to cross the halo is smaller than our simulation time-step. 
As we described in Section~\ref{subsec:numerical_setup_binary_single}, we use 10 spherical shells only for the $> O(10^{7}) \, M_{\odot}$ DM halos.

\begin{figure}[ht!]
        \centering
        \includegraphics[width=\linewidth]{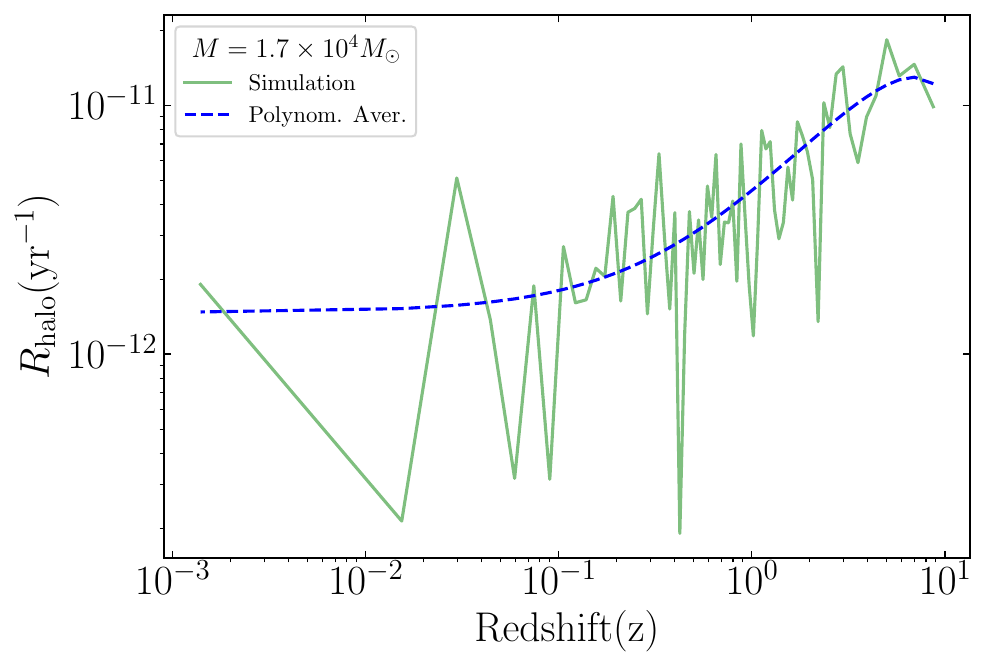}\\

        \includegraphics[width=\linewidth]{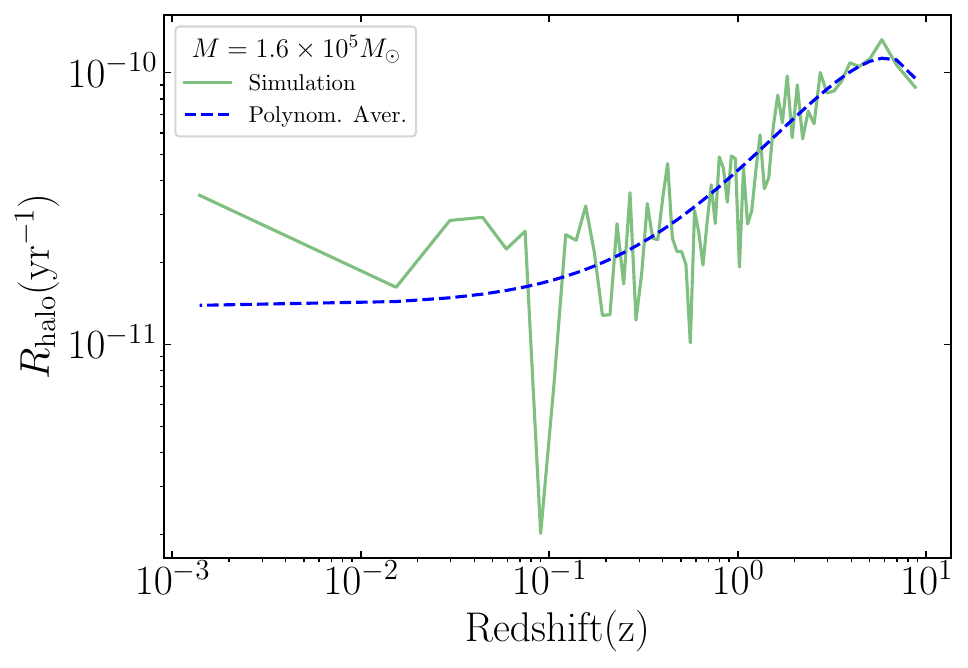} \\
        \includegraphics[width=\linewidth]{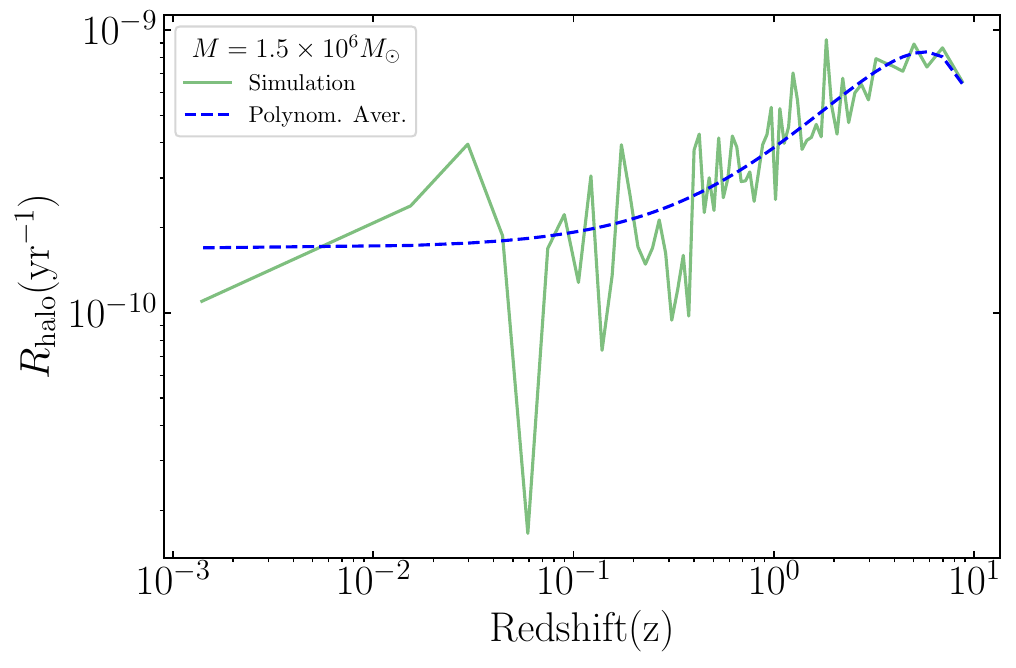}
    \vspace{-0.2cm}
    \caption{\justifying As in Fig.~\ref{3body_rate_1e3}, for a halo with mass $M=1.7\times 10^{4} M_{\odot}$ (top),
    $M=1.6\times 10^{6}M_{\odot}$ (middle) and
    $M=1.5\times 10^{6}M_{\odot}$ (bottom)
    at $z=0$. Two, three, and five spherical shells have been used for each halo, respectively.}
    \label{3body_rate_1e4_5_6}
\end{figure}

We see a similar trend in halos of all masses. In Fig.~\ref{3body_rate_1e4_5_6}, we present the merger rate of more massive halos. As the halo mass increases, the merger rate increases as well. 
In fact, the merger rate per halo for masses from $10^{3}$ to $10^{8} \, M_{\odot}$ increases approximately proportionally to the mass of the DM halo. 
That is unlike the direct capture rates where the merger rate per halo increases more slowly with mass. 

As shown in Fig.~\ref{N_merger_cum_v1}, for the $10^{12} \, M_{\odot}$ halo, there are many PBH mergers in the inner shells before re-scaling, but few in the outer ones. 
For these massive halos, after re-scaling our simulations, the many mergers in the inner shells have a small contribution to the total merger rate, while the few merger events in the outer shells dominate the rate at any given time. 
The stochastic nature of these few merger events in the outer shells causes the merger rates to fluctuate significantly between time-steps. This is resolved by using appropriate polynomial averaging.   
  
\begin{figure}[h!t]
    \centering
    \includegraphics[width=\linewidth]{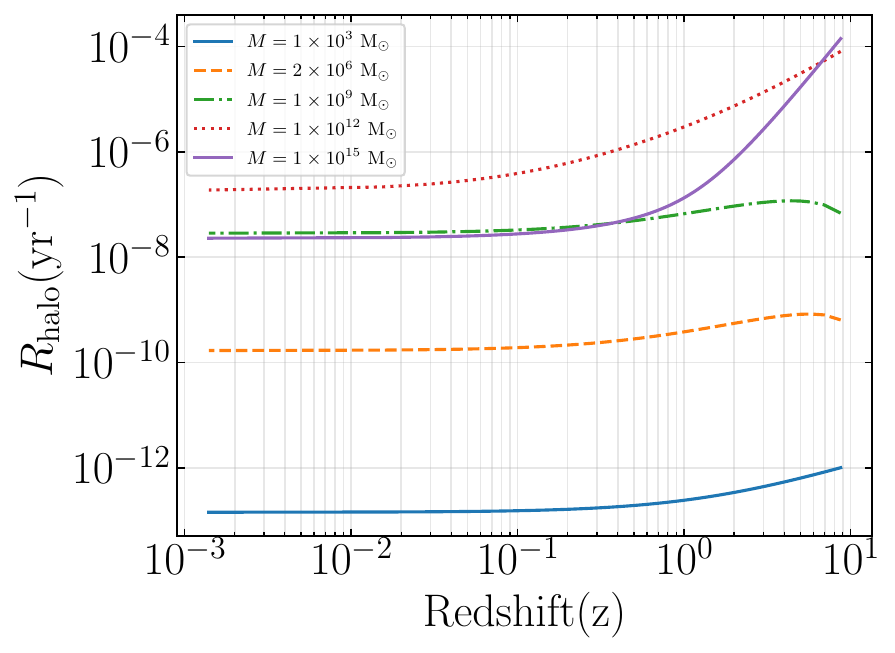}
    \caption{\justifying Evolution of the merger rate per halo $R_{\textrm{halo}}(z)$, from the binary-single interactions. We show results for halos of different masses at $z=0$. The \textit{Ludlow16} concentration model has been used, incorporating a monochromatic PBH mass-distribution with $M_{\textrm{PBH}}=30 \, M_{\odot}$.}
    \label{3body_rate_halos}
\end{figure}

In Fig.~\ref{3body_rate_halos}, we compare the merger rate per halo for several halo masses. 
We notice that the merger rate per halo from binary-single interactions is much higher than the one from two-body captures presented in Fig.~\ref{R_halo_evol_masses}. This occurs because the binary-single interactions dominate when the combination of $f_{\textrm{PBH}}$ and $f_{\textrm{PBH binaries}}$ is high, as we have taken it to be here. 
Hence, the environment is dense in PBH binaries and in PBHs in general. 
We also note that as the halo mass increases, the merger rate per halo increases and faster than it does with the direct capture case (compare to Fig.~\ref{R_halo_evol_masses}). 
Given the \textit{Press-Schechter} halo mass function (and most conventional halo mass functions), for the rate due to binary-single interactions a much wider range of DM halo masses contribute. This makes these rates less susceptible to arguments about the stability of the smallest halo masses. 
At $z=0$, it is the halos with mass from $10^{3}-10^{9} M_{\odot}$ that contribute to the PBH merger rate due to binary-single interactions. 
By comparison for the direct capture rates only the DM halos of the smallest masses dominate the total rate.
Finally, for some of the most massive -galaxy cluster scale- DM halos, even the rate per halo gets suppressed.
In these halos, the relative velocities between the PBH binaries and the surrounding single PBHs are too high for any hardening to take place. 

\begin{figure}[ht!]
    \centering
    \includegraphics[width=\linewidth]{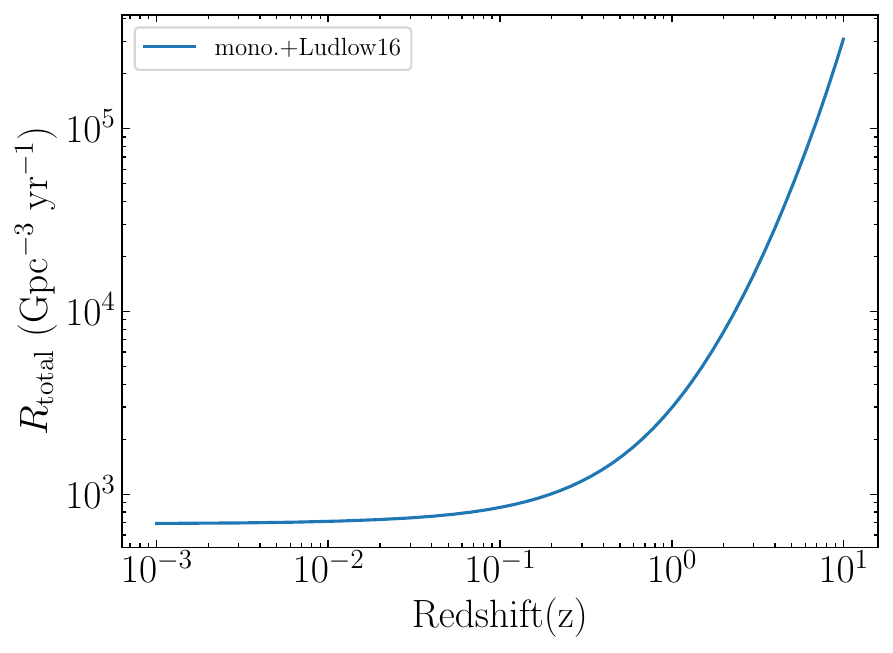} 
    \caption{\justifying {The redshift evolution of the PBH merger rate from binary-single interactions, after summing over all halo masses. We have taken $f_{\textrm{single \, PBH}}=1/2$, $f_{\textrm{PBH \, binaries}}=1/2$ and used the \textit{Press-Schechter} halo mass function with PBH mass $m=30 \, M_{\odot}$. We plot up to redshift of 10.}}
    \label{R_total_PS}
\end{figure}

The total comoving merger rate due to binary-single interactions is presented in Fig.~\ref{R_total_PS} where a set of 50 DM halos with masses logarithmically spaced between $10^3$ and $10^{15}$ $M_\odot$ are simulated to the present day. We used a monochromatic PBH mass distribution and employed \textit{Ludlow16} concentration model for each halo. The \textit{Press-Schechter} halo mass function is utilized in the calculation of the total rate. For the choice of $f_{\textrm{PBH}} = 1$ and $f_{\textrm{single PBH}} = f_{\textrm{PBH binaries}} = 1/2$, the expected merger rate from binary-single interactions is much larger than total merger rate from two-body capture presented in Fig.~\ref{R2b_total_PS}. 

\begin{figure}[ht!]
    \centering
    \includegraphics[width=0.9\linewidth]{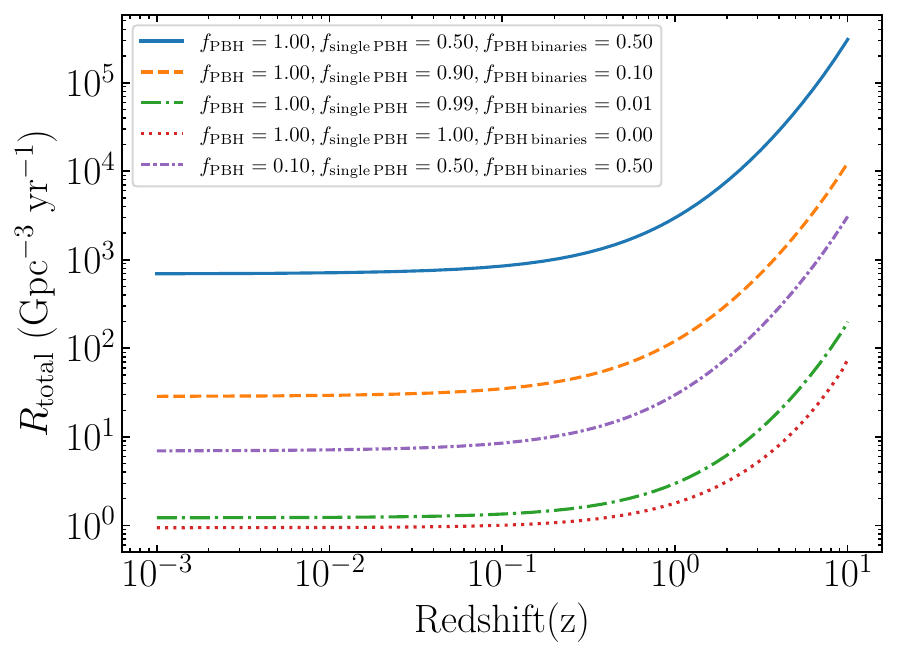} 
    \caption{\justifying Total comoving merger rate $R_{\textrm{total}}$ versus redshift $z$ for different PBHs fractions. The total rate includes contributions from two-body captures and binary-single interactions. The five lines correspond to various choices of $f_{\textrm{PBH \, binaries}}/f_{\textrm{single \, PBH}}$ and on $f_{\textrm{PBH}}$. We plot up to redshift of 10.}
    \label{R_total_fractions}
\end{figure}

Even if all of DM is in PBHs at the stellar mass range ($f_{\textrm{PBH}} = 1$), the relative ratio of single PBHs to PBH binaries ($f_{\textrm{PBH \, binaries}}/f_{\textrm{single \, PBH}}$) is highly uncertain. 
In Fig.~\ref{R_total_fractions}, we show the total comoving PBH merger rate as a function of redshift, from the contribution of direct capture events and from binary-single interactions. 
We show five lines for that rate for alternative assumptions on $f_{\textrm{PBH}}$ and $f_{\textrm{PBH \, binaries}}$. 
We always assume that $f_{\textrm{single \, PBH}} + f_{\textrm{PBH \, binaries}} = 1$. 
For $f_{\textrm{PBH \, binaries}}/f_{\textrm{single \, PBH}} \sim 1$ the total rate is dominated by the binary single interaction rate. Compare the solid blue line of Fig.~\ref{R_total_fractions} to Fig.~\ref{R_total_PS}, they are effectively identical. Only once $f_{\textrm{PBH \, binaries}}/f_{\textrm{single \, PBH}}$ is $O(10^{-2})$ do the two components become comparable (green dashed-dotted line). 
The red dotted line is for the case where only the direct capture rate is present.

We examined the impact of a halo mass function with suppressed small-scale power, as proposed by \textit{Jenkins} \cite{Jenkins}. Up to $z \approx 1$, the \textit{Press-Schechter} function predicts merger rates 6-8 times higher than \textit{Jenkins}' model. Above $z\approx 1$, both models increase sharply, with \textit{Press-Schechter} still leading as the gap narrows.

The total PBH merger rate  scales as $f_{\textrm{PBH}}^{2}$ which means that any choice of $f_{PBH}$ can be rescaled accordingly. Thus, lower values of $f_{PBH}$ will reduce rates but do not fundamentally change our qualitative conclusions.
We note that these rates compared to the rates of \cite{PBBH1, Franciolini:2022ewd, Martinelli:2022elq} that study the impact of unperturbed binaries merging at redshift of 0 are about a factor of $10^{4}$ smaller (at $z=0$). Revisiting the assumption on the rate of mergers from the unperturbed PBH binaries is left for a future study.

\section{Conclusions}
\label{sec:conclusions}

Primordial black holes remain a viable candidate for the explanation of the observed abundance of DM. 
In this work we revisit the merger rate of PBH pairs with masses in the stellar mass range that can be observed by the ground based gravitational wave observatories. We focus on i) the hard binaries that can form in direct capture events, when two PBHs inside DM halos come close enough to emit in gravitational waves enough of their initial energy to form a bound boundary and ii) the interactions between PBH binaries that have existed as bound to each other since their formation, at the early universe and once they fall in a DM halo can interact multiple times with single PBHs, accelerating the evolution of their orbital properties, leading to merger events.

In this work, we use the halo mass accretion history to calculate the merger rates at each redshift, both for the two-body captures and for binary-single interactions. 
We also use the information on the concentration parameter of DM halos, 
that is relevant to model the evolving density of PBHs inside halos and the evolution with redshift of their velocity dispersion profiles. 
We also take into account that for the massive halos at any given time, there is a radial profile on their velocity dispersion. 
This allows us to evaluate the merger rate as a function of the redshift going back as far as $z=10$, an era that has not been explored in that context. 
Examples of these rates for given DM halos are given in Figs.~\ref{fig:R_halo_evol} and~\ref{R_halo_evol_masses}, for the binaries formed via direct capture interactions and Figs.\ref{3body_rate_1e3},~\ref{3body_rate_1e4_5_6} and~\ref{3body_rate_halos}, for the rates associated with binaries hardening by interacting with their environment.

We find that the PBH (comoving) merger rates increase dramatically as we go back in time with our results shown in Fig.~\ref{R2b_total_PS} for the direct capture rates and Fig.~\ref{R_total_PS} for the rate associated to the binary-single interactions.
While the direct capture rate is at all times dominated by the contribution of the smallest DM halos available, the rate due to the binary-single interactions receives a significant contribution from a wide range of DM halo masses. 
At $z=0$ that includes masses as large as $10^{9} \, M_{\odot}$.  
In reality both rates can be present. Depending on the fraction of DM in stellar mass black holes, and the relative abundance of single PBHs vs PBHs in binaries the total rate can be dominated by either component as we show in Fig.~\ref{R_total_fractions}.  
However, even if $O(10^{-2})$ of the PBHs are formed in binaries that survive falling into a DM halo, we expect the total merger rate to have an important if not dominant contribution from the PBH binary-single PBH interactions taking place inside DM halos.
Ref.~\cite{Kritos} focusing just on small-mass DM halos found an appreciable contribution to the PBH merger rate from binary-single interactions. We find that our results are compatible with theirs.
We also study the typical time-scale for hardening interactions of PBH binaries with their environment. We find that the main point of Ref.~\cite{early_PBH_1} is true under the conditions that they focus on. However, we note again the importance of mid-sized DM halos in the total PBH merger rate due to binary-single interactions in them. Even if rare, the mergers from that mechanism are likely more common than the mergers due to direct captures.

We study alternative PBH mass distributions, finding that as long as there is a peak of these distributions in the stellar mass range the total merger rates are fairly similar to those evaluated for a monochromatic mass distribution. 
Given the current limits on PBHs from gravitational microlensing \cite{Blaineau:2022nhy, Mroz:2024mse} (see however \cite{Garcia-Bellido:2024yaz}), there is little motivation to study a PBH mass distribution heavily tilted on the subsolar-mass ranges. 
Such a distribution would need separate treatment to account for the fact that the smaller PBHs could be ejected entirely out of the host DM halo, an effect which we ignore due to the narrow PBH mass distribution that we study.

Given the high merger rates at early times, future ground based observatories as the Einstein Telescope \cite{Hild:2010id} and the cosmic explorer \cite{Evans:2023euw, Gupta:2023lga}, will be able to probe high redshifts and directly probe these PBH-PBH binaries even if a small fraction of DM is in the stellar-mass range.

\begin{acknowledgments}
We thank Konstantinos Kritos for valuable discussions at the early stages of this work. MA thanks to Steven Murray for his guidance on using the HMFcalc package. This material is based upon work supported by the National Science Foundation, under Grant$\#$2207912. 
\end{acknowledgments}

\begin{appendix}

\section{Two-body capture rate per halo assuming a PBH mass distribution}
\label{App:PBH_mass_distr_2_body}

In this appendix, we present the calculation of the two-body capture rate for a generalized PBH mass distribution. 
The differential capture rate of PBHs with masses $m_1$ and $m_2$ is
\begin{equation}
\label{Rate_mass_functioj}
\Gamma_{\textrm{capture}, m_1, m_2}
 = 4\pi  \int_{0}^{R_{\textrm{vir}}} \langle \sigma \cdot v_{\textrm{pbh}} \rangle \cdot n_{m_{1}}\cdot n_{m_{2}} \cdot r^2 \, dr.
\end{equation}
The $n_{m_2}$ and $n_{m_1}$ are the number densities of PBHs with mass $m_1$ and $m_2$. They are equal to,
\begin{equation}\label{N_m1}
n_{m_1}=\frac{\rho_{NFW}(r) \cdot f_{\textrm{ PBH}} \cdot f_{m_{1}}}{m_1}
\end{equation}
and
\begin{equation}\label{N_m2}
n_{m_2}=\frac{\rho_{NFW}(r) \cdot f_{\textrm{PBH}} \cdot f_{m_{2}}}{m_2}.
\end{equation}
where $f_{m_1}$ and $f_{m_2}$ indicate the fractions of PBHs with masses $m_1$ and $m_2$, respectively.

Substituting Eqs.~\ref{N_m1} and  \ref{N_m2} into (\ref{Rate_mass_functioj}) we get,
\begin{eqnarray}
\Gamma_{\textrm{capture}, m_1, m_2} &=& \frac{4\pi \cdot f_{\textrm{ PBH}}^2 \cdot f_{m_{1}} \cdot f_{m_{2}}}{m_{1} \cdot m_{2}} \\ 
&\times&  \int_{0}^{R_{\textrm{vir}}} \langle \sigma \cdot v_{\textrm{pbh}} \rangle \cdot \rho_{NFW}(r)^2 \cdot r^2 \, dr. \nonumber
\end{eqnarray}

To obtain the rate, we first integrate the radial part over the volume, taking the limit of $r$ to $R_{vir}$,
\begin{eqnarray}
\int_0^{R_{\textrm{vir}}} dr \, r^2 \rho_{\textrm{NFW}}^2(r) &=& \frac{1}{3} R_s^6 \rho_s^2 \left[\frac{1}{R_s^3} - \frac{1}{(R_{\textrm{vir}}+R_s)^3}\right] \nonumber \\
&=& \frac{1}{3} R_s^3 \rho_s^2 f(C).
\label{integral}
\end{eqnarray}

Here, 
\begin{equation}
f(C)=\left[1 - \frac{1}{(1+C)^3}\right].
\end{equation}

Substituting (\ref{integral}) into (\ref{Rate_mass_functioj}) we get,
\begin{eqnarray}
\label{Rate_mass_fucntion_2}
\Gamma_{\textrm{capture}, m_1, m_2} &=& \frac{1 }{12\pi}\frac{ f_{\textrm{PBH}}^2 \cdot f_{m1} f_{m2}}{m_1 m_2}  \nonumber \\
&\times& \frac{M_{\textrm{vir}}^2 f(C)}{ R_s^3 g(c)^2}   \cdot \langle \sigma \cdot v_{pbh}\rangle.
\label{Rate_mass_fucntion_2}
\end{eqnarray}

The average of $\sigma \cdot v_{pbh}$ is given by,

\begin{eqnarray}
\left\langle\sigma \cdot v_{\textrm{pbh}}\right\rangle &=& \int d^3 v_{\textrm{pbh}}  \cdot v_{\textrm{pbh}} \cdot p(v_{\textrm{pbh}}) \cdot \sigma(v_{\textrm{pbh}}) \\
&=& 4\pi \int_0^{v_{\textrm{vir}}} dv_{\textrm{pbh}} \, v_{\textrm{pbh}}^{3} \, p(v_{\textrm{pbh}}) \cdot \sigma(v_{\textrm{pbh}}). \nonumber
\label{eq:average}
\end{eqnarray}
Here $p(v_{\textrm{pbh}})$ is the probability density function for $v_{\textrm{pbh}}$.
Substituting Eq.(\ref{eq:cross-section}) into Eq.(\ref{eq:average})

\begin{eqnarray}
\label{Cross_section_final}
\left\langle\sigma \cdot v_{pbh}\right\rangle &=& 2\pi\left(\frac{85\pi}{6\sqrt{2}}\right)^{2/7} \nonumber \\
&\times& \frac{G^2(m_1+m_2)^{10/7}m_1^{2/7}m_2^{2/7}}{c^{10/7}} \nonumber \\
&\times& 4\pi \int_0^{v_{\textrm{vir}}}dv_{\textrm{pbh}}v_{\textrm{pbh}}^{3/7}p(v_{pbh}). 
\end{eqnarray}
Substituting (\ref{Cross_section_final}) into (\ref{Rate_mass_fucntion_2}) we obtain,
\begin{eqnarray}
\Gamma_{\textrm{capture}, m_1, m_2}&=&  \frac{1 }{12\pi}\frac{ f_{\textrm{PBH}}^2 \cdot f_{m_1} f_{m_1}}{m_1 m_2} \frac{M_{\textrm{vir}}^2}{ R_s^3 g(c)^2} f(C) \nonumber \\
&\times&  2\pi\left(\frac{85\pi}{6\sqrt{2}}\right)^{2/7} \nonumber \\
&\times& \frac{G^2(m_1+m_2)^{10/7}m_1^{2/7}m_2^{2/7}}{c^{10/7}}  \nonumber \\
&\times& 4\pi \int_0^{v_{\textrm{vir}}}dv_{\textrm{pbh}}v_{\textrm{pbh}}^{3/7} p(v_{pbh}). 
\end{eqnarray}

\begin{eqnarray}
\Gamma_{\textrm{capture}, m_1, m_2} &=& 
\frac{2 \pi}{3}\left(\frac{85\pi}{6\sqrt{2}}\right)^{2/7} \frac{G^2 \cdot f_{\textrm{PBH}}^2 \cdot f_{m_1} f_{m_2}}{c} 
 \nonumber \\
&\times& \frac{(m_1+m_2)^{10/7}m_1^{2/7}m_2^{2/7}}{m_1 m_2} \cdot 
\nonumber \\
&\times& \left(\frac{M_{\textrm{vir}}^2}{R_s^3 g(c)^2}f(C)\right) \cdot D(v)
\end{eqnarray}

\begin{equation}
D(v)=\int_0^{v_{v i r}} p(v_{pbh})\left(\frac{ v_{pbh}}{c}\right)^{3 / 7} d v.
\end{equation}
The capture rate provided above applies when single PBHs  in the DM halo have specific masses $m_1$ and $m_2$. However, when the masses follow a distribution it is necessary to incorporate a probability distribution for PBH masses into the rate that accounts for the various probabilities of different mass pairs and ensures a correct calculation of capture rates across the halo. In this regard, we use a generic PBH mass distribution, $\psi(m_{1})$ and $\psi(m_{2})$, (discussed further in Appendix~\ref{PBH_PDFs}) and integrate it over a reasonable mass range of maximum and minimum values. Our new capture  rate, $\Gamma_{\textrm{capture}}$ then becomes,
\begin{eqnarray}
\Gamma_{\textrm{capture}} &=&\frac{1}{2}\int_{m_{1}^{\textrm{min}}}^{m_{1}^{\textrm{max}}} \int_{m_{2}^{\textrm{min}}}^{m_{2}^{\textrm{max}}} \frac{2\pi}{3} \left(\frac{85 \pi}{6 \sqrt{2}}\right)^{2 / 7} \frac{G^2 \cdot f_{\textrm{PBH}}^2  }{c} \nonumber \\
&\times& \frac{(m_1+m_2)^{10 / 7} m_1^{2 / 7} m_2^{2 / 7}}{ \, m_1 \, m_2} \cdot \psi(m_1) \cdot \psi(m_2) 
 \nonumber \\
&\times&
\frac{M_{\textrm{vir}}^2}{R_s^3 g(c)^2} f(C) \cdot D(v)  dm_1  dm_2.
\end{eqnarray}
Note that we have omitted the fractions $f_{m_1}$ and $f_{m_2}$ since we no longer assume that individual PBHs have monochromatic mass but follow a mass distribution. Moreover, $1/2$ is imposed because we want to avoid double counting of unique pairs.
The final form of the merger rate per halo then is,
\begin{eqnarray}
\Gamma_{\textrm{capture}} &=&K\int_{m_{1}^{\textrm{min}}}^{m_{1}^{\textrm{max}}} \int_{m_{2}^{\textrm{min}}}^{m_{2}^{\textrm{max}}} 
\frac{(m_1+m_2)^{10 / 7} m_1^{2 / 7} m_2^{2 / 7}}{ \, m_1 \, m_2}  \nonumber \\
&\times&  \psi(m_1) \cdot \psi(m_2)  dm_1   dm_2  \nonumber \\
&\times& \frac{M_{\textrm{vir}}^2}{R_s^3 g(c)^2} \cdot f(c)\cdot D(v),
\end{eqnarray}
with 
\begin{eqnarray}
K &=& \frac{1}{2}\cdot \frac{2\pi}{3}\left(\frac{85 \pi}{6 \sqrt{2}}\right)^{2 / 7} \frac{G^2 f_{\textrm{ \, PBH}}^2  }{c}, \nonumber \\
f(C) &=& 1 - \frac{1}{(1+C)^3}, \\
D(v) &=& \int_0^{v_{\textrm{vir}}} P\left(v, v_{\textrm{dm}}\right)\left(\frac{v}{c}\right)^{3 / 7} \, dv. \nonumber
\end{eqnarray}

The above rate includes  all the possible captures between single PBHs with different masses in any given halo. This is covered by the integration over $m_1$ and $m_2$  from 5 $M_\odot$ to 150 $M_\odot$, covering all PBH masses within this specified range.

\section{PBHs Mass Functions}
\label{PBH_PDFs}
We use three different PBH mass distributions commonly referenced in the literature. These mass distributions include the log-normal, power-law, broken power-law, and critical collapse (CC) distributions presented in \cite{Chen:2024dxh}. We normalize the probability all three distributions such that 
\begin{equation}
\int \psi(dm) \, dm = 1
\end{equation}with the integration taken over the mass range \(5 \, M_\odot\) to \(150 \, M_\odot\). 

\subsubsection{Lognormal Mass distribution} 
The log-normal mass distribution is defined by,
\begin{equation}
\psi(m)=\frac{1}{\sqrt{2 \pi} \sigma m} \exp \left(-\frac{\left(\ln(m) -\mu\right)^2}{2 \sigma^2}\right), 
\end{equation}
with $\mu =ln(M_c)$  where  $M_c$ represents the median mass and $\sigma$ characterizes the width of the mass distribution.
In Fig.~\ref{fig:PBH_PDF}, we have presented a graph of this distribution with a median mean of $\mu=\ln(30 M_\odot)$ and different values of variance $\sigma$.

\subsubsection{Broken Power-law Mass distribution} 
The broken power law mass distribution is given by,
\begin{equation}
\psi(m) = \left[\frac{m_*}{\alpha_1 + 1} + \frac{m_*}{\alpha_2 - 1}\right]^{-1} 
\begin{cases} 
\left(\frac{m}{m_*}\right)^{\alpha_1}, & \text{if } m < m_* \\ 
\left(\frac{m}{m_*}\right)^{-\alpha_2}, & \text{if } m > m_*
\end{cases}
\end{equation}

where $m_*$ is the peak mass and $\alpha_1>0$, $\alpha_2>1$ are power-law parameters. We take the best-fit values given in Ref. \cite{Chen:2024dxh} which are $m_*$=31.1, $\alpha_1$=0.54 and $\alpha_2$=5.6. An illustration of this mass distribution is presented in the top panel of Fig. \ref{fig:PBH_PDF} for reference.

\subsubsection{Critical Collapse Mass distribution}
\begin{figure}[!htb]
    \centering
    \begin{minipage}[b]{0.9\linewidth}
        \centering
        \includegraphics[width=\linewidth]{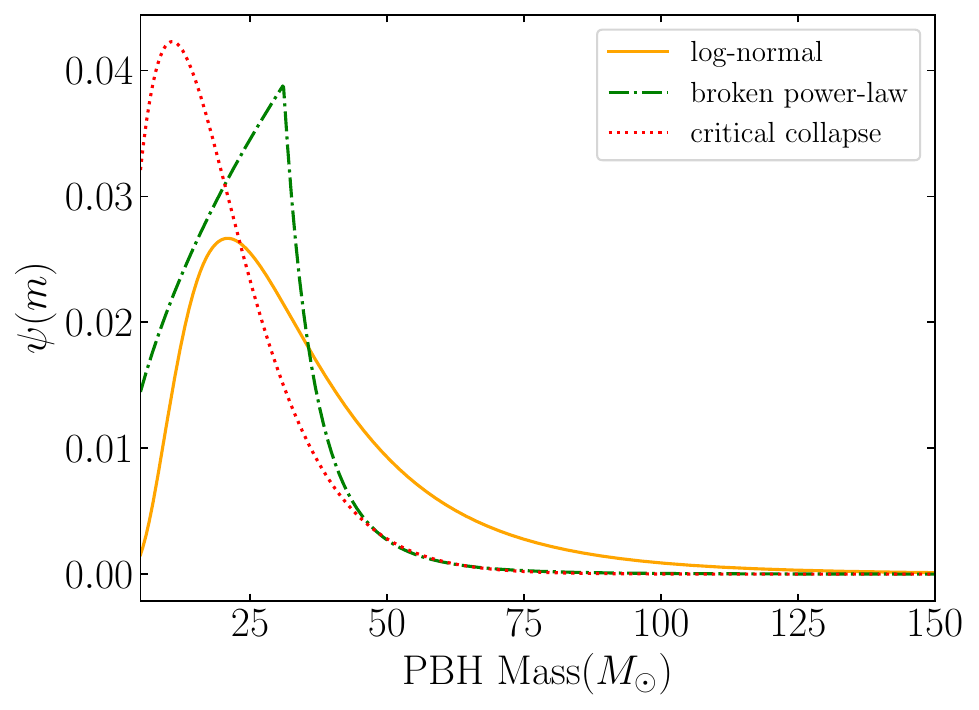}
        \label{}
    \end{minipage}
    \vfill
    \begin{minipage}[b]{0.9\linewidth}
        \centering
        \includegraphics[width=\linewidth]{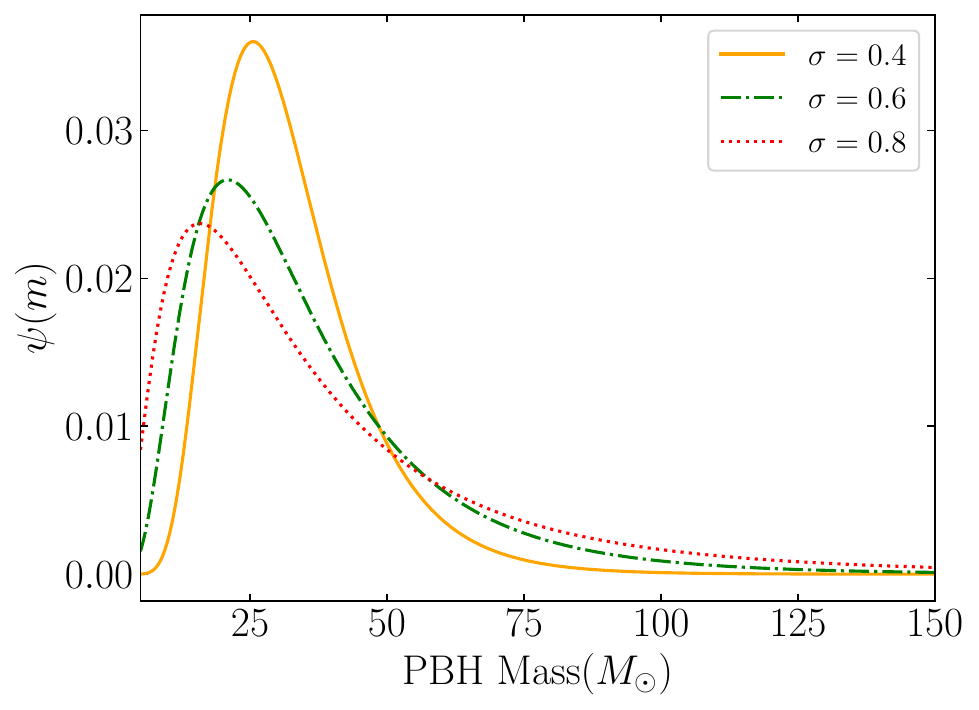}
        \label{}
    \end{minipage}
    \caption{\justifying Mass distributions for PBHs: log-normal  (with $\mu = \ln(30M_\odot)$ and $\sigma = 0.71$), broken power-law  (with $M_{*}=31.1 M_\odot$, $\alpha=0.54$, and $\alpha_{2}=5.6$), and the critical collapse  distribution (with $M_{f}=10.8M_\odot$ and $\alpha=1.1$) (\textit{top}). Log-normal distributions with different variance $\sigma$ are shown in the \textit{bottom} panel. All the distributions are normalized over the range of $5-150$ $M_{\odot}$.}
    \label{fig:PBH_PDF}
\end{figure}

The critical collapse mass arises from the critical collapse of radiation that leads to the formation of PBHs. It is characterized by,
\begin{equation}
\psi(m)=\frac{\alpha^2 m^\alpha}{M_{\textrm{f}}^{1+\alpha} \Gamma_{\textrm{}}(1 / \alpha)} \exp \left(-\left(m / M_{\textrm{f}}\right)^\alpha\right), 
\end{equation}
where $\alpha$ is a universal exponent related to the critical collapse, and $M_{\textrm{f}}$ is a mass scale approximately of the order of the horizon mass at the collapse epoch. This distribution experiences exponential suppression beyond the mass scale of $M_{\textrm{f}}$. We take the best fit values given in Ref \cite{Chen:2024dxh} which are $M_{\textrm{f}}=10.8 M_{\odot}$ and $\alpha=1$.  An illustration of this mass distribution is depict in the top panel of Fig. \ref{fig:PBH_PDF} for reference.

\section{Mass accretion history} \label{MAH_params}
The parameters $\alpha$ and $\beta$ in Eq.(\ref{eq:MAH}) govern the mass accretion history of halos. Each halo with mass $M$ today has unique values of these parameters that govern the mass accretion history of the halos over time. To determine these parameters, we need to know the formation redshift $ z_{-2} $ of the halo which is defined as redshift when the total mass of the halo $M(z)$ is equal to the enclosed mass within the scale scale $R_s$, $M_{r}(<R_s)$ and its concentration $C$ at present.  For each DM its formation time $ z_{-2} $, can be calculated using cosmology-dependent constants, with and the parameters $\alpha$ and $\beta$ subsequently derived \cite{MAH},
\begin{eqnarray}
z_{-2} &=& \left( \frac{200}{A_{cosmo}} \frac{C(M_0,z_{0})^3 g(1)}{\Omega_{\textrm{m}} g(C(M_0))} - \frac{\Omega_{\Lambda}}{\Omega_{\textrm{m}}} \right)^{1/3}- 1\\
\alpha &=& \frac{\ln(g(1) / g(C)) - \beta z_{-2}}{\ln(1+z_{-2})} \;  \\
\beta &=& -\frac{3}{1+z_{-2}},
\label{eq:Mass_accr_param}
\end{eqnarray}
where $A_{cosmo}=798$.We employ \textit{Ludlow16} model for the mass-concentration-redshift relation of $C$ in our simulation.

\section{PBH binaries' initial orbital parameters distribution at formation}
\label{app:Prestime_PBHbinaries}
To obtain the initial distribution of orbital parameters of the PBH binaries, we start by defining the mean PBH separation at matter-radiation equality as,
\begin{equation}
    \bar{x} =\left(\frac{3 m_{\textrm{PBH}}}{4 \pi f_{\textrm{PBH}} \rho_{\textrm{eq}}}\right)^{1 / 3}.
\end{equation}
Here $\rho_{\textrm {eq}}$ is the average energy density in the matter-radiation equality ($z\approx3450$). PBHs follow a Poisson spatial distribution at formation and their differential probability distribution of the re-scaled angular momentum $j \equiv \sqrt{1-e^2}$ reads \cite{PBBH1,PBBH2,PBBH3,PBBH4},
\begin{eqnarray}
 P(j) &=&\frac{y(j)^2}{j\left(1+y(j)^2\right)^{3 / 2}}, \\
y(j) & = & \frac{j}{0.5\left(1+\sigma_{\textrm{eq}}^2 / f_{\textrm{PBH}}^2\right)^{1 / 2}(x / \bar{x})^3}.
\label{eq:P_j}
\end{eqnarray}
Here $\sigma_{\textrm{eq}} \approx 0.005$ represents the variance of the Gaussian large-scale density fluctuations at the epoch of matter-radiation equality. This distribution results from the combined effect of nearby PBHs and matter perturbations that exert a torque on the binary PBH system during its formation \cite{PBBH1}. Ultimately, the distribution that characterizes both $j$ and the semi-major axis $a$ can be expressed as,
\begin{equation}\label{P_a_j}
P(a, j)=\frac{3  \cdot a^{-1 / 4}}{4}\left(\frac{f_{\textrm{PBH}}}{\alpha \bar{x}}\right)^{3 / 4} P(j) \exp \left[-\left(\frac{x(a)}{\bar{x}}\right)^3\right],
\end{equation}
with
\begin{equation}
x(a) =\left(\frac{3 a m_{\textrm{PBH}}}{4 \pi \alpha \rho_{\textrm{eq}}}\right)^ {1 / 4},
\end{equation}
and  $\alpha=0.1$.

\begin{figure}[h!]
    \centering
    \begin{minipage}{0.9\linewidth}
        \centering
        \includegraphics[width=\linewidth]{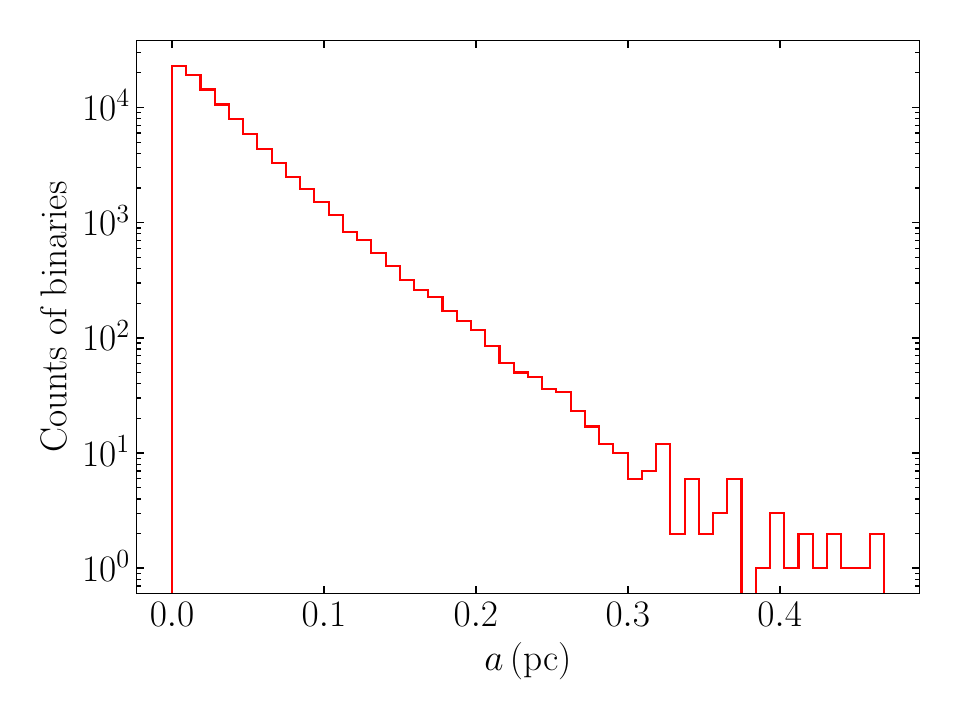}
        \label{fig:subfig1}
    \end{minipage}\hfill
    \begin{minipage}{0.9\linewidth}
        \centering
        \includegraphics[width=\linewidth]{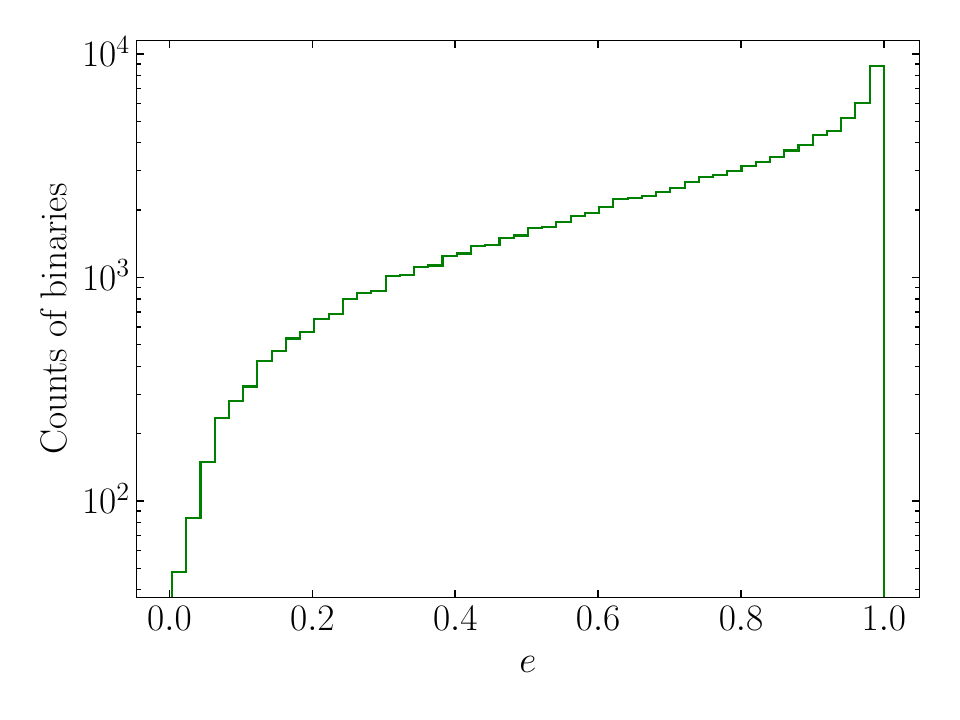}
        \label{fig:subfig2}
    \end{minipage}
    \caption{\justifying The initial distribution of semi-major axis(\textit{top}) and eccentricity (\textit{bottom})of PBH binaries for $N=10^5$ binaries. We assume all PBHs have mass of $m=30\, M_\odot$.}
    \label{fig:PBBHs_a_e}
\end{figure}

In Fig.~\ref{fig:PBBHs_a_e}, we show the counts for initial distribution of eccentricity $e$ and semi-major axis $a$ for PBH binaries formed at matter-radiation equality. 
First, we sampled the re-scaled angular momentum $j$ from its probability distribution $ P(j) $, ensuring that $ j $ values range between (0, 1). Then, we used the joint distribution $ P(a, j) $ to sample $ a_0 $ by randomly selecting values within ($ 10^{-6}$ , 1) pc. In both cases, we employ the inverse sampling techniques with $ N_{\textrm{sample}} = 10^4 $ to generate samples. We also assumed that $f_{\textrm{BH}} = 1$ and $m_{\textrm{BH}}=30M_\odot$ each. From the plot, it is clear that PBH binaries with high eccentricities have the highest counts and small semi-major axes. The mean value, $a_{mean}$, of the semi-major axis from the pristine distributions in in Eq.(\ref{P_a_j}) is  around 7287 AU.

In  table \ref{N_hard}  we also show  the percentage of hard binaries at redshift $z=0$ in different  halos. The shell number indicates the index of the shells in each halo, with lower indices indicating the inner shells of a halo and higher indices representing the outer shells. It is obvious that that low-mass halos possess a greater proportion of hard binaries, where all binaries are characterized as hard. moreover, massive halos contain fewer hard binaries. Furthermore, within a specific halo, more hard binaries are present in the inner shells compared to the outer shells. These are the type of binaries that we evolve and analyze in our simulations are merging within the Hubble time.

\begin{table}[h!]
\centering
\begin{tabular}{|c|c|c|c|}
    \hline
           Halo Mass & No of shells & $a_{hard}$(au) & $N_{hard} ( \%)$  \\
        \hline
        $ 10^{3}$ & - & $4.48 \times 10^{5}$ & 100.000 \\
                \hline

        $ 10^{6}$ & 1 & $2.08 \times 10^{-1}$ & 99.634 \\
                \hline

        $ 10^{6}$ & 3 & $1.79 \times 10^{-2}$ & 41.183 \\
                \hline

        $ 10^{6}$ & 5 & $3.17 \times 10^{-2}$ & 61.199 \\
                \hline

        $ 10^{9}$ & 1 & $7.07 \times 10^{-2}$ & 87.365 \\
                \hline

        $ 10^{9}$ & 6 & $4.20 \times 10^{-4}$ & 0.670 \\
                \hline

        $ 10^{9}$ & 10 & $3.17 \times 10^{-4}$ & 0.461 \\
                \hline

        $ 10^{12}$ & 1 & $1.07 \times 10^{-2}$ & 26.523 \\
                \hline

        $ 10^{12}$ & 6 & $1.79 \times 10^{-5}$ & 0.022 \\
                \hline

        $ 10^{12}$ & 10 & $3.17 \times 10^{-6}$ & 0.003 \\
                \hline

        $ 10^{15}$ & 1 & $2.26 \times 10^{-3}$ & 5.130 \\
                \hline

        $ 10^{15}$ & 6 & $1.21 \times 10^{-6}$ & 0.000 \\
                \hline

        $ 10^{15}$ & 10 & $3.17 \times 10^{-8}$ & 0.000 \\
                \hline

        \end{tabular}
    \caption{The percentage of hard binaries ,$N_{hard} (\%)$,in different halos  and shells at $z=0$}
    \label{N_hard}
\end{table}

\section{Binary-single interaction time scale of PBH binaries}\label{Timescale_app}
In this Appendix, we present the timescale for the binary-single interaction during the evolution of a given halo. 
For a binary with a semi-major axis of $a=a_h$ to interact with a third PBH, and hence the rate interactions between a PBH binary and a third PBH is given by
\begin{equation}
R_{3b}=2 \pi G \frac{m_T \cdot  n(r,z) \cdot a}{v_{\textrm{disp}}(r,z)},
\end{equation}
where $m_{T}$ = $m_\textrm{PBH \, binaries}+m_{\textrm{single \, PBH}}$, $n(r,t)$ is the number density of the PBHs in the environment where the BBH is located, and $v_{\textrm{disp}}(r,z)$ is the velocity dispersion of the  single PBHs. In the case of monochromatic mass, $m_{T}=3m$. We define the number density of single PBHs as $n(r,t)=\frac{\rho_{\textrm{env}}(r,z)}{m}$, which is basically $\rho_{\textrm{env}}(r,z) =\rho_{\textrm{NFW}}(r,z)$. The final form of the rate for $a=a_h$ is then, 
\begin{equation}
R_{3b}=6 \pi G \frac{   \rho_{\textrm{NFW}}(r,z)\cdot a_{h}}{v_{\textrm{disp}},(r,z)}
\end{equation}
which is equivalent to the  timescale of, 
\begin{equation} 
\tau_{3b}=\frac{v_{\textrm{disp}}(r,z)}{6 \pi G \cdot  \rho_{\textrm{NFW}},(r,z)\cdot a_{h}}.
\end{equation}

We have presented the value of this timescale in Table \ref{timescales} for different DM mass halos and at different redshift
in the evolution of those haloes. We notice that binary-single hard interactions become significant at a late stage in their evolution as the timescales get smaller.

\begin{table}[h!]
\centering
\begin{tabular}{|c|c|c|c|c|c|}
\hline
\text{M($z$) ($M_\odot$)} & z & $R_{i}$ & \text{No. of shells} & $a_{h}$ (\text{au}) & $\tau_{3b}$ (\text{Myr}) \\

\hline
$1.0 \times 10^{3}$ & 0 & $R_{\textrm{vir}/2}$ & 1 & $4.5 \times 10^{5}$ & $5.5 \times 10^{4}$ \\
$6.2 \times 10^{2}$ & 2 & $R_{\textrm{vir}/2}$ & 1 & $3.3 \times 10^{5}$ & $8.1 \times 10^{3}$ \\
$1.2 \times 10^{2}$ & 8 & $R_{\textrm{vir}/2}$ & 1 & $4.1 \times 10^{5}$ & $1.9 \times 10^{2}$ \\
\hline
$1.0 \times 10^{6}$ & 0 & 1 & 5 & $4.3 \times 10^{4}$ & $2.6 \times 10^{1}$ \\
$1.0 \times 10^{6}$ & 0 & 3 & 5 & $3.7 \times 10^{3}$ & $7.9 \times 10^{4}$ \\
$1.0 \times 10^{6}$ & 0 & 5 & 5 & $6.5 \times 10^{3}$ & $5.1 \times 10^{7}$ \\
$5.6 \times 10^{5}$ & 2 & 1 & 5 & $4.5 \times 10^{4}$ & $1.6 \times 10^{1}$ \\
$5.6 \times 10^{5}$ & 2 & 3 & 5 & $4.3 \times 10^{3}$ & $2.5 \times 10^{4}$ \\
$5.6 \times 10^{5}$ & 2 & 5 & 5 & $4.6 \times 10^{3}$ & $8.5 \times 10^{6}$ \\
$7.3 \times 10^{4}$ & 8 & 1 & 5 & $5.8 \times 10^{4}$ & $5.1 \times 10^{0}$ \\
$7.3 \times 10^{4}$ & 8 & 3 & 5 & $8.6 \times 10^{3}$ & $2.2 \times 10^{3}$ \\
$7.3 \times 10^{4}$ & 8 & 5 & 5 & $6.2 \times 10^{3}$ & $2.2 \times 10^{5}$ \\
\hline
$1.0 \times 10^{9}$ & 0 & 1 & 10 & $1.5 \times 10^{4}$ & $1.0 \times 10^{1}$ \\
$1.0 \times 10^{9}$ & 0 & 6 & 10 & $8.7 \times 10^{1}$ & $9.7 \times 10^{6}$ \\
$1.0 \times 10^{9}$ & 0 & 10 & 10 & $6.5 \times 10^{1}$ & $6.5 \times 10^{10}$ \\
$4.6 \times 10^{8}$ & 2 & 1 & 10 & $1.4 \times 10^{4}$ & $7.6 \times 10^{0}$ \\
$4.6 \times 10^{8}$ & 2 & 6 & 10 & $1.2 \times 10^{2}$ & $2.6 \times 10^{6}$ \\
$4.6 \times 10^{8}$ & 2 & 10 & 10 & $5.3 \times 10^{1}$ & $8.7 \times 10^{9}$ \\
$3.1 \times 10^{7}$ & 8 & 1 & 10 & $1.5 \times 10^{4}$ & $3.6 \times 10^{0}$ \\
$3.1 \times 10^{7}$ & 8 & 6 & 10 & $2.9 \times 10^{2}$ & $1.6 \times 10^{5}$ \\
$3.1 \times 10^{7}$ & 8 & 10 & 10 & $1.1 \times 10^{2}$ & $1.1 \times 10^{8}$ \\
\hline
$1.0 \times 10^{12}$ & 0 & 1 & 10 & $2.2 \times 10^{3}$ & $5.2 \times 10^{1}$ \\
$1.0 \times 10^{12}$ & 0 & 6 & 10 & $3.7 \times 10^{0}$ & $6.9 \times 10^{8}$ \\
$1.0 \times 10^{12}$ & 0 & 10 & 10 & $6.5 \times 10^{-1}$ & $4.7 \times 10^{13}$ \\
$2.9 \times 10^{11}$ & 2 & 1 & 10 & $1.9 \times 10^{3}$ & $4.0 \times 10^{1}$ \\
$2.9 \times 10^{11}$ & 2 & 6 & 10 & $5.2 \times 10^{0}$ & $1.5 \times 10^{8}$ \\
$2.9 \times 10^{11}$ & 2 & 10 & 10 & $7.2 \times 10^{-1}$ & $4.2 \times 10^{12}$ \\
$4.3 \times 10^{9}$ & 8 & 1 & 10 & $2.4 \times 10^{3}$ & $1.6 \times 10^{1}$ \\
$4.3 \times 10^{9}$ & 8 & 6 & 10 & $2.1 \times 10^{1}$ & $4.1 \times 10^{6}$ \\
$4.3 \times 10^{9}$ & 8 & 10 & 10 & $4.1 \times 10^{0}$ & $1.3 \times 10^{10}$ \\
\hline
$1.0 \times 10^{15}$ & 0 & 1 & 10 & $4.7 \times 10^{2}$ & $2.1 \times 10^{2}$ \\
$1.0 \times 10^{15}$ & 0 & 6 & 10 & $2.5 \times 10^{-1}$ & $4.0 \times 10^{10}$ \\
$1.0 \times 10^{15}$ & 0 & 10 & 10 & $6.5 \times 10^{-3}$ & $3.4 \times 10^{16}$ \\
$1.6 \times 10^{14}$ & 2 & 1 & 10 & $2.9 \times 10^{2}$ & $1.9 \times 10^{2}$ \\
$1.6 \times 10^{14}$ & 2 & 6 & 10 & $3.0 \times 10^{-1}$ & $7.6 \times 10^{9}$ \\
$1.6 \times 10^{14}$ & 2 & 10 & 10 & $1.1 \times 10^{-2}$ & $1.9 \times 10^{15}$ \\
$2.9 \times 10^{11}$ & 8 & 1 & 10 & $5.4 \times 10^{2}$ & $5.5 \times 10^{1}$ \\
$2.9 \times 10^{11}$ & 8 & 6 & 10 & $2.4 \times 10^{0}$ & $6.4 \times 10^{7}$ \\
$2.9 \times 10^{11}$ & 8 & 10 & 10 & $2.4 \times 10^{-1}$ & $7.9 \times 10^{11}$ \\
\hline
\end{tabular}
    \caption{\justifying{Binary-single interaction timescale for a binary with semi-major axis $a = a_h$ to interact with PBH singles at different redshifts ($z = 0$, $2$, $8$), located at different spherical shells for halos with masses of $ 10^3$,  $10^6$, $10^9$, $ 10^{12}$, and $10^{15}$ M$_\odot$ at $z = 0$.}}

\label{timescales}
\end{table}
\end{appendix}
\bibliography{paper}

\begin{thebibliography}{50}%
\makeatletter
\providecommand \@ifxundefined [1]{%
 \@ifx{#1\undefined}
}%
\providecommand \@ifnum [1]{%
 \ifnum #1\expandafter \@firstoftwo
 \else \expandafter \@secondoftwo
 \fi
}%
\providecommand \@ifx [1]{%
 \ifx #1\expandafter \@firstoftwo
 \else \expandafter \@secondoftwo
 \fi
}%
\providecommand \natexlab [1]{#1}%
\providecommand \enquote  [1]{``#1''}%
\providecommand \bibnamefont  [1]{#1}%
\providecommand \bibfnamefont [1]{#1}%
\providecommand \citenamefont [1]{#1}%
\providecommand \href@noop [0]{\@secondoftwo}%
\providecommand \href [0]{\begingroup \@sanitize@url \@href}%
\providecommand \@href[1]{\@@startlink{#1}\@@href}%
\providecommand \@@href[1]{\endgroup#1\@@endlink}%
\providecommand \@sanitize@url [0]{\catcode `\\12\catcode `\$12\catcode `\&12\catcode `\#12\catcode `\^12\catcode `\_12\catcode `\%12\relax}%
\providecommand \@@startlink[1]{}%
\providecommand \@@endlink[0]{}%
\providecommand \url  [0]{\begingroup\@sanitize@url \@url }%
\providecommand \@url [1]{\endgroup\@href {#1}{\urlprefix }}%
\providecommand \urlprefix  [0]{URL }%
\providecommand \Eprint [0]{\href }%
\providecommand \doibase [0]{https://doi.org/}%
\providecommand \selectlanguage [0]{\@gobble}%
\providecommand \bibinfo  [0]{\@secondoftwo}%
\providecommand \bibfield  [0]{\@secondoftwo}%
\providecommand \translation [1]{[#1]}%
\providecommand \BibitemOpen [0]{}%
\providecommand \bibitemStop [0]{}%
\providecommand \bibitemNoStop [0]{.\EOS\space}%
\providecommand \EOS [0]{\spacefactor3000\relax}%
\providecommand \BibitemShut  [1]{\csname bibitem#1\endcsname}%
\let\auto@bib@innerbib\@empty
\bibitem [{\citenamefont {Zel'dovich}\ and\ \citenamefont {Novikov}(1967)}]{PBH_1}%
  \BibitemOpen
  \bibfield  {author} {\bibinfo {author} {\bibfnamefont {Y.~B.}\ \bibnamefont {Zel'dovich}}\ and\ \bibinfo {author} {\bibfnamefont {I.~D.}\ \bibnamefont {Novikov}},\ }\bibfield  {title} {\bibinfo {title} {{The Hypothesis of Cores Retarded during Expansion and the Hot Cosmological Model}},\ }\href@noop {} {\bibfield  {journal} {\bibinfo  {journal} {Sov. Astron.}\ }\textbf {\bibinfo {volume} {10}},\ \bibinfo {pages} {602} (\bibinfo {year} {1967})}\BibitemShut {NoStop}%
\bibitem [{\citenamefont {{Chapline}}(1975)}]{PBH_2}%
  \BibitemOpen
  \bibfield  {author} {\bibinfo {author} {\bibfnamefont {G.~F.}\ \bibnamefont {{Chapline}}},\ }\bibfield  {title} {\bibinfo {title} {{Cosmological effects of primordial black holes}},\ }\href {https://doi.org/10.1038/253251a0} {\bibfield  {journal} {\bibinfo  {journal} {\nat}\ }\textbf {\bibinfo {volume} {253}},\ \bibinfo {pages} {251} (\bibinfo {year} {1975})}\BibitemShut {NoStop}%
\bibitem [{\citenamefont {{Hawking}}(1971)}]{PBH_3}%
  \BibitemOpen
  \bibfield  {author} {\bibinfo {author} {\bibfnamefont {S.}~\bibnamefont {{Hawking}}},\ }\bibfield  {title} {\bibinfo {title} {{Gravitationally collapsed objects of very low mass}},\ }\href {https://doi.org/10.1093/mnras/152.1.75} {\bibfield  {journal} {\bibinfo  {journal} {\mnras}\ }\textbf {\bibinfo {volume} {152}},\ \bibinfo {pages} {75} (\bibinfo {year} {1971})}\BibitemShut {NoStop}%
\bibitem [{\citenamefont {Carr}\ \emph {et~al.}(2016)\citenamefont {Carr}, \citenamefont {Kuhnel},\ and\ \citenamefont {Sandstad}}]{PBH_4}%
  \BibitemOpen
  \bibfield  {author} {\bibinfo {author} {\bibfnamefont {B.}~\bibnamefont {Carr}}, \bibinfo {author} {\bibfnamefont {F.}~\bibnamefont {Kuhnel}},\ and\ \bibinfo {author} {\bibfnamefont {M.}~\bibnamefont {Sandstad}},\ }\bibfield  {title} {\bibinfo {title} {{Primordial Black Holes as Dark Matter}},\ }\href {https://doi.org/10.1103/PhysRevD.94.083504} {\bibfield  {journal} {\bibinfo  {journal} {Phys. Rev. D}\ }\textbf {\bibinfo {volume} {94}},\ \bibinfo {pages} {083504} (\bibinfo {year} {2016})},\ \Eprint {https://arxiv.org/abs/1607.06077} {arXiv:1607.06077 [astro-ph.CO]} \BibitemShut {NoStop}%
\bibitem [{\citenamefont {Abbott}\ \emph {et~al.}(2016)\citenamefont {Abbott} \emph {et~al.}}]{LIGOScientific:2016aoc}%
  \BibitemOpen
  \bibfield  {author} {\bibinfo {author} {\bibfnamefont {B.~P.}\ \bibnamefont {Abbott}} \emph {et~al.} (\bibinfo {collaboration} {LIGO Scientific, Virgo}),\ }\bibfield  {title} {\bibinfo {title} {{Observation of Gravitational Waves from a Binary Black Hole Merger}},\ }\href {https://doi.org/10.1103/PhysRevLett.116.061102} {\bibfield  {journal} {\bibinfo  {journal} {Phys. Rev. Lett.}\ }\textbf {\bibinfo {volume} {116}},\ \bibinfo {pages} {061102} (\bibinfo {year} {2016})},\ \Eprint {https://arxiv.org/abs/1602.03837} {arXiv:1602.03837 [gr-qc]} \BibitemShut {NoStop}%
\bibitem [{\citenamefont {{Bird}}\ \emph {et~al.}(2016)\citenamefont {{Bird}}, \citenamefont {{Cholis}}, \citenamefont {{Mu{\~n}oz}}, \citenamefont {{Ali-Ha{\"\i}moud}}, \citenamefont {{Kamionkowski}}, \citenamefont {{Kovetz}}, \citenamefont {{Raccanelli}},\ and\ \citenamefont {{Riess}}}]{Bird:2016dcv}%
  \BibitemOpen
  \bibfield  {author} {\bibinfo {author} {\bibfnamefont {S.}~\bibnamefont {{Bird}}}, \bibinfo {author} {\bibfnamefont {I.}~\bibnamefont {{Cholis}}}, \bibinfo {author} {\bibfnamefont {J.~B.}\ \bibnamefont {{Mu{\~n}oz}}}, \bibinfo {author} {\bibfnamefont {Y.}~\bibnamefont {{Ali-Ha{\"\i}moud}}}, \bibinfo {author} {\bibfnamefont {M.}~\bibnamefont {{Kamionkowski}}}, \bibinfo {author} {\bibfnamefont {E.~D.}\ \bibnamefont {{Kovetz}}}, \bibinfo {author} {\bibfnamefont {A.}~\bibnamefont {{Raccanelli}}},\ and\ \bibinfo {author} {\bibfnamefont {A.~G.}\ \bibnamefont {{Riess}}},\ }\bibfield  {title} {\bibinfo {title} {{Did LIGO Detect Dark Matter?}},\ }\href {https://doi.org/10.1103/PhysRevLett.116.201301} {\bibfield  {journal} {\bibinfo  {journal} {\prl}\ }\textbf {\bibinfo {volume} {116}},\ \bibinfo {eid} {201301} (\bibinfo {year} {2016})},\ \Eprint {https://arxiv.org/abs/1603.00464} {arXiv:1603.00464 [astro-ph.CO]} \BibitemShut {NoStop}%
\bibitem [{\citenamefont {Carr}\ \emph {et~al.}(2021)\citenamefont {Carr}, \citenamefont {Kohri}, \citenamefont {Sendouda},\ and\ \citenamefont {Yokoyama}}]{PBH_DM_1}%
  \BibitemOpen
  \bibfield  {author} {\bibinfo {author} {\bibfnamefont {B.}~\bibnamefont {Carr}}, \bibinfo {author} {\bibfnamefont {K.}~\bibnamefont {Kohri}}, \bibinfo {author} {\bibfnamefont {Y.}~\bibnamefont {Sendouda}},\ and\ \bibinfo {author} {\bibfnamefont {J.}~\bibnamefont {Yokoyama}},\ }\bibfield  {title} {\bibinfo {title} {{Constraints on primordial black holes}},\ }\href {https://doi.org/10.1088/1361-6633/ac1e31} {\bibfield  {journal} {\bibinfo  {journal} {Rept. Prog. Phys.}\ }\textbf {\bibinfo {volume} {84}},\ \bibinfo {pages} {116902} (\bibinfo {year} {2021})},\ \Eprint {https://arxiv.org/abs/2002.12778} {arXiv:2002.12778 [astro-ph.CO]} \BibitemShut {NoStop}%
\bibitem [{\citenamefont {Carr}\ and\ \citenamefont {Kuhnel}(2020)}]{PBH_DM_2}%
  \BibitemOpen
  \bibfield  {author} {\bibinfo {author} {\bibfnamefont {B.}~\bibnamefont {Carr}}\ and\ \bibinfo {author} {\bibfnamefont {F.}~\bibnamefont {Kuhnel}},\ }\bibfield  {title} {\bibinfo {title} {{Primordial Black Holes as Dark Matter: Recent Developments}},\ }\href {https://doi.org/10.1146/annurev-nucl-050520-125911} {\bibfield  {journal} {\bibinfo  {journal} {Ann. Rev. Nucl. Part. Sci.}\ }\textbf {\bibinfo {volume} {70}},\ \bibinfo {pages} {355} (\bibinfo {year} {2020})},\ \Eprint {https://arxiv.org/abs/2006.02838} {arXiv:2006.02838 [astro-ph.CO]} \BibitemShut {NoStop}%
\bibitem [{\citenamefont {Green}\ and\ \citenamefont {Kavanagh}(2021)}]{PBH_DM_3}%
  \BibitemOpen
  \bibfield  {author} {\bibinfo {author} {\bibfnamefont {A.~M.}\ \bibnamefont {Green}}\ and\ \bibinfo {author} {\bibfnamefont {B.~J.}\ \bibnamefont {Kavanagh}},\ }\bibfield  {title} {\bibinfo {title} {{Primordial Black Holes as a dark matter candidate}},\ }\href {https://doi.org/10.1088/1361-6471/abc534} {\bibfield  {journal} {\bibinfo  {journal} {J. Phys. G}\ }\textbf {\bibinfo {volume} {48}},\ \bibinfo {pages} {043001} (\bibinfo {year} {2021})},\ \Eprint {https://arxiv.org/abs/2007.10722} {arXiv:2007.10722 [astro-ph.CO]} \BibitemShut {NoStop}%
\bibitem [{\citenamefont {Kovetz}\ \emph {et~al.}(2017)\citenamefont {Kovetz}, \citenamefont {Cholis}, \citenamefont {Breysse},\ and\ \citenamefont {Kamionkowski}}]{Kovetz:2016kpi}%
  \BibitemOpen
  \bibfield  {author} {\bibinfo {author} {\bibfnamefont {E.~D.}\ \bibnamefont {Kovetz}}, \bibinfo {author} {\bibfnamefont {I.}~\bibnamefont {Cholis}}, \bibinfo {author} {\bibfnamefont {P.~C.}\ \bibnamefont {Breysse}},\ and\ \bibinfo {author} {\bibfnamefont {M.}~\bibnamefont {Kamionkowski}},\ }\bibfield  {title} {\bibinfo {title} {{Black hole mass function from gravitational wave measurements}},\ }\href {https://doi.org/10.1103/PhysRevD.95.103010} {\bibfield  {journal} {\bibinfo  {journal} {Phys. Rev. D}\ }\textbf {\bibinfo {volume} {95}},\ \bibinfo {pages} {103010} (\bibinfo {year} {2017})},\ \Eprint {https://arxiv.org/abs/1611.01157} {arXiv:1611.01157 [astro-ph.CO]} \BibitemShut {NoStop}%
\bibitem [{\citenamefont {Clesse}\ and\ \citenamefont {Garcia-Bellido}(2022)}]{Clesse:2020ghq}%
  \BibitemOpen
  \bibfield  {author} {\bibinfo {author} {\bibfnamefont {S.}~\bibnamefont {Clesse}}\ and\ \bibinfo {author} {\bibfnamefont {J.}~\bibnamefont {Garcia-Bellido}},\ }\bibfield  {title} {\bibinfo {title} {{GW190425, GW190521 and GW190814: Three candidate mergers of primordial black holes from the QCD epoch}},\ }\href {https://doi.org/10.1016/j.dark.2022.101111} {\bibfield  {journal} {\bibinfo  {journal} {Phys. Dark Univ.}\ }\textbf {\bibinfo {volume} {38}},\ \bibinfo {pages} {101111} (\bibinfo {year} {2022})},\ \Eprint {https://arxiv.org/abs/2007.06481} {arXiv:2007.06481 [astro-ph.CO]} \BibitemShut {NoStop}%
\bibitem [{\citenamefont {Clesse}\ \emph {et~al.}(2018)\citenamefont {Clesse}, \citenamefont {Garc\'\i{}a-Bellido},\ and\ \citenamefont {Orani}}]{Clesse:2018ogk}%
  \BibitemOpen
  \bibfield  {author} {\bibinfo {author} {\bibfnamefont {S.}~\bibnamefont {Clesse}}, \bibinfo {author} {\bibfnamefont {J.}~\bibnamefont {Garc\'\i{}a-Bellido}},\ and\ \bibinfo {author} {\bibfnamefont {S.}~\bibnamefont {Orani}},\ }\bibfield  {title} {\bibinfo {title} {{Detecting the Stochastic Gravitational Wave Background from Primordial Black Hole Formation}},\ }\href@noop {} {\  (\bibinfo {year} {2018})},\ \Eprint {https://arxiv.org/abs/1812.11011} {arXiv:1812.11011 [astro-ph.CO]} \BibitemShut {NoStop}%
\bibitem [{\citenamefont {Sasaki}\ \emph {et~al.}(2016)\citenamefont {Sasaki}, \citenamefont {Suyama}, \citenamefont {Tanaka},\ and\ \citenamefont {Yokoyama}}]{early_PBH_1}%
  \BibitemOpen
  \bibfield  {author} {\bibinfo {author} {\bibfnamefont {M.}~\bibnamefont {Sasaki}}, \bibinfo {author} {\bibfnamefont {T.}~\bibnamefont {Suyama}}, \bibinfo {author} {\bibfnamefont {T.}~\bibnamefont {Tanaka}},\ and\ \bibinfo {author} {\bibfnamefont {S.}~\bibnamefont {Yokoyama}},\ }\bibfield  {title} {\bibinfo {title} {{Primordial Black Hole Scenario for the Gravitational-Wave Event GW150914}},\ }\href {https://doi.org/10.1103/PhysRevLett.117.061101} {\bibfield  {journal} {\bibinfo  {journal} {Phys. Rev. Lett.}\ }\textbf {\bibinfo {volume} {117}},\ \bibinfo {pages} {061101} (\bibinfo {year} {2016})},\ \bibinfo {note} {[Erratum: Phys.Rev.Lett. 121, 059901 (2018)]},\ \Eprint {https://arxiv.org/abs/1603.08338} {arXiv:1603.08338 [astro-ph.CO]} \BibitemShut {NoStop}%
\bibitem [{\citenamefont {Ali-Ha\"\i{}moud}\ \emph {et~al.}(2017)\citenamefont {Ali-Ha\"\i{}moud}, \citenamefont {Kovetz},\ and\ \citenamefont {Kamionkowski}}]{PBBH1}%
  \BibitemOpen
  \bibfield  {author} {\bibinfo {author} {\bibfnamefont {Y.}~\bibnamefont {Ali-Ha\"\i{}moud}}, \bibinfo {author} {\bibfnamefont {E.~D.}\ \bibnamefont {Kovetz}},\ and\ \bibinfo {author} {\bibfnamefont {M.}~\bibnamefont {Kamionkowski}},\ }\bibfield  {title} {\bibinfo {title} {{Merger rate of primordial black-hole binaries}},\ }\href {https://doi.org/10.1103/PhysRevD.96.123523} {\bibfield  {journal} {\bibinfo  {journal} {Phys. Rev. D}\ }\textbf {\bibinfo {volume} {96}},\ \bibinfo {pages} {123523} (\bibinfo {year} {2017})},\ \Eprint {https://arxiv.org/abs/1709.06576} {arXiv:1709.06576 [astro-ph.CO]} \BibitemShut {NoStop}%
\bibitem [{\citenamefont {Kovetz}(2017)}]{Kovetz:2017rvv}%
  \BibitemOpen
  \bibfield  {author} {\bibinfo {author} {\bibfnamefont {E.~D.}\ \bibnamefont {Kovetz}},\ }\bibfield  {title} {\bibinfo {title} {{Probing Primordial-Black-Hole Dark Matter with Gravitational Waves}},\ }\href {https://doi.org/10.1103/PhysRevLett.119.131301} {\bibfield  {journal} {\bibinfo  {journal} {Phys. Rev. Lett.}\ }\textbf {\bibinfo {volume} {119}},\ \bibinfo {pages} {131301} (\bibinfo {year} {2017})},\ \Eprint {https://arxiv.org/abs/1705.09182} {arXiv:1705.09182 [astro-ph.CO]} \BibitemShut {NoStop}%
\bibitem [{\citenamefont {Morras}\ \emph {et~al.}(2023)\citenamefont {Morras} \emph {et~al.}}]{Morras:2023jvb}%
  \BibitemOpen
  \bibfield  {author} {\bibinfo {author} {\bibfnamefont {G.}~\bibnamefont {Morras}} \emph {et~al.},\ }\bibfield  {title} {\bibinfo {title} {{Analysis of a subsolar-mass compact binary candidate from the second observing run of Advanced LIGO}},\ }\href {https://doi.org/10.1016/j.dark.2023.101285} {\bibfield  {journal} {\bibinfo  {journal} {Phys. Dark Univ.}\ }\textbf {\bibinfo {volume} {42}},\ \bibinfo {pages} {101285} (\bibinfo {year} {2023})},\ \Eprint {https://arxiv.org/abs/2301.11619} {arXiv:2301.11619 [gr-qc]} \BibitemShut {NoStop}%
\bibitem [{\citenamefont {Abbott}\ \emph {et~al.}(2023)\citenamefont {Abbott} \emph {et~al.}}]{LIGOScientific:2022hai}%
  \BibitemOpen
  \bibfield  {author} {\bibinfo {author} {\bibfnamefont {R.}~\bibnamefont {Abbott}} \emph {et~al.} (\bibinfo {collaboration} {LIGO Scientific, VIRGO, KAGRA}),\ }\bibfield  {title} {\bibinfo {title} {{Search for subsolar-mass black hole binaries in the second part of Advanced LIGO's and Advanced Virgo's third observing run}},\ }\href {https://doi.org/10.1093/mnras/stad588} {\bibfield  {journal} {\bibinfo  {journal} {Mon. Not. Roy. Astron. Soc.}\ }\textbf {\bibinfo {volume} {524}},\ \bibinfo {pages} {5984} (\bibinfo {year} {2023})},\ \bibinfo {note} {[Erratum: Mon.Not.Roy.Astron.Soc. 526, 6234 (2023)]},\ \Eprint {https://arxiv.org/abs/2212.01477} {arXiv:2212.01477 [astro-ph.HE]} \BibitemShut {NoStop}%
\bibitem [{\citenamefont {Kawasaki}\ \emph {et~al.}(2012)\citenamefont {Kawasaki}, \citenamefont {Kusenko},\ and\ \citenamefont {Yanagida}}]{PBH_super_1}%
  \BibitemOpen
  \bibfield  {author} {\bibinfo {author} {\bibfnamefont {M.}~\bibnamefont {Kawasaki}}, \bibinfo {author} {\bibfnamefont {A.}~\bibnamefont {Kusenko}},\ and\ \bibinfo {author} {\bibfnamefont {T.~T.}\ \bibnamefont {Yanagida}},\ }\bibfield  {title} {\bibinfo {title} {{Primordial seeds of supermassive black holes}},\ }\href {https://doi.org/10.1016/j.physletb.2012.03.056} {\bibfield  {journal} {\bibinfo  {journal} {Phys. Lett. B}\ }\textbf {\bibinfo {volume} {711}},\ \bibinfo {pages} {1} (\bibinfo {year} {2012})},\ \Eprint {https://arxiv.org/abs/1202.3848} {arXiv:1202.3848 [astro-ph.CO]} \BibitemShut {NoStop}%
\bibitem [{\citenamefont {Nakama}\ \emph {et~al.}(2016)\citenamefont {Nakama}, \citenamefont {Suyama},\ and\ \citenamefont {Yokoyama}}]{PBH_super_2}%
  \BibitemOpen
  \bibfield  {author} {\bibinfo {author} {\bibfnamefont {T.}~\bibnamefont {Nakama}}, \bibinfo {author} {\bibfnamefont {T.}~\bibnamefont {Suyama}},\ and\ \bibinfo {author} {\bibfnamefont {J.}~\bibnamefont {Yokoyama}},\ }\bibfield  {title} {\bibinfo {title} {{Supermassive black holes formed by direct collapse of inflationary perturbations}},\ }\href {https://doi.org/10.1103/PhysRevD.94.103522} {\bibfield  {journal} {\bibinfo  {journal} {Phys. Rev. D}\ }\textbf {\bibinfo {volume} {94}},\ \bibinfo {pages} {103522} (\bibinfo {year} {2016})},\ \Eprint {https://arxiv.org/abs/1609.02245} {arXiv:1609.02245 [gr-qc]} \BibitemShut {NoStop}%
\bibitem [{\citenamefont {Hasegawa}\ and\ \citenamefont {Kawasaki}(2018)}]{PBH_super_3}%
  \BibitemOpen
  \bibfield  {author} {\bibinfo {author} {\bibfnamefont {F.}~\bibnamefont {Hasegawa}}\ and\ \bibinfo {author} {\bibfnamefont {M.}~\bibnamefont {Kawasaki}},\ }\bibfield  {title} {\bibinfo {title} {{Cogenesis of LIGO Primordial Black Holes and Dark Matter}},\ }\href {https://doi.org/10.1103/PhysRevD.98.043514} {\bibfield  {journal} {\bibinfo  {journal} {Phys. Rev. D}\ }\textbf {\bibinfo {volume} {98}},\ \bibinfo {pages} {043514} (\bibinfo {year} {2018})},\ \Eprint {https://arxiv.org/abs/1711.00990} {arXiv:1711.00990 [astro-ph.CO]} \BibitemShut {NoStop}%
\bibitem [{\citenamefont {Kawasaki}\ and\ \citenamefont {Murai}(2019)}]{PBH_super_4}%
  \BibitemOpen
  \bibfield  {author} {\bibinfo {author} {\bibfnamefont {M.}~\bibnamefont {Kawasaki}}\ and\ \bibinfo {author} {\bibfnamefont {K.}~\bibnamefont {Murai}},\ }\bibfield  {title} {\bibinfo {title} {{Formation of supermassive primordial black holes by Affleck-Dine mechanism}},\ }\href {https://doi.org/10.1103/PhysRevD.100.103521} {\bibfield  {journal} {\bibinfo  {journal} {Phys. Rev. D}\ }\textbf {\bibinfo {volume} {100}},\ \bibinfo {pages} {103521} (\bibinfo {year} {2019})},\ \Eprint {https://arxiv.org/abs/1907.02273} {arXiv:1907.02273 [astro-ph.CO]} \BibitemShut {NoStop}%
\bibitem [{\citenamefont {Raidal}\ \emph {et~al.}(2024)\citenamefont {Raidal}, \citenamefont {Vaskonen},\ and\ \citenamefont {Veerm\"ae}}]{PBH_channels}%
  \BibitemOpen
  \bibfield  {author} {\bibinfo {author} {\bibfnamefont {M.}~\bibnamefont {Raidal}}, \bibinfo {author} {\bibfnamefont {V.}~\bibnamefont {Vaskonen}},\ and\ \bibinfo {author} {\bibfnamefont {H.}~\bibnamefont {Veerm\"ae}},\ }\bibfield  {title} {\bibinfo {title} {{Formation of primordial black hole binaries and their merger rates}},\ }\href@noop {} {\  (\bibinfo {year} {2024})},\ \Eprint {https://arxiv.org/abs/2404.08416} {arXiv:2404.08416 [astro-ph.CO]} \BibitemShut {NoStop}%
\bibitem [{\citenamefont {Mandic}\ \emph {et~al.}(2016)\citenamefont {Mandic}, \citenamefont {Bird},\ and\ \citenamefont {Cholis}}]{Mandic}%
  \BibitemOpen
  \bibfield  {author} {\bibinfo {author} {\bibfnamefont {V.}~\bibnamefont {Mandic}}, \bibinfo {author} {\bibfnamefont {S.}~\bibnamefont {Bird}},\ and\ \bibinfo {author} {\bibfnamefont {I.}~\bibnamefont {Cholis}},\ }\bibfield  {title} {\bibinfo {title} {{Stochastic Gravitational-Wave Background due to Primordial Binary Black Hole Mergers}},\ }\href {https://doi.org/10.1103/PhysRevLett.117.201102} {\bibfield  {journal} {\bibinfo  {journal} {Phys. Rev. Lett.}\ }\textbf {\bibinfo {volume} {117}},\ \bibinfo {pages} {201102} (\bibinfo {year} {2016})},\ \Eprint {https://arxiv.org/abs/1608.06699} {arXiv:1608.06699 [astro-ph.CO]} \BibitemShut {NoStop}%
\bibitem [{\citenamefont {{Cholis}}\ \emph {et~al.}(2016)\citenamefont {{Cholis}}, \citenamefont {{Kovetz}}, \citenamefont {{Ali-Ha{\"\i}moud}}, \citenamefont {{Bird}}, \citenamefont {{Kamionkowski}}, \citenamefont {{Mu{\~n}oz}},\ and\ \citenamefont {{Raccanelli}}}]{Cholis:2016kqi}%
  \BibitemOpen
  \bibfield  {author} {\bibinfo {author} {\bibfnamefont {I.}~\bibnamefont {{Cholis}}}, \bibinfo {author} {\bibfnamefont {E.~D.}\ \bibnamefont {{Kovetz}}}, \bibinfo {author} {\bibfnamefont {Y.}~\bibnamefont {{Ali-Ha{\"\i}moud}}}, \bibinfo {author} {\bibfnamefont {S.}~\bibnamefont {{Bird}}}, \bibinfo {author} {\bibfnamefont {M.}~\bibnamefont {{Kamionkowski}}}, \bibinfo {author} {\bibfnamefont {J.~B.}\ \bibnamefont {{Mu{\~n}oz}}},\ and\ \bibinfo {author} {\bibfnamefont {A.}~\bibnamefont {{Raccanelli}}},\ }\bibfield  {title} {\bibinfo {title} {{Orbital eccentricities in primordial black hole binaries}},\ }\href {https://doi.org/10.1103/PhysRevD.94.084013} {\bibfield  {journal} {\bibinfo  {journal} {\prd}\ }\textbf {\bibinfo {volume} {94}},\ \bibinfo {eid} {084013} (\bibinfo {year} {2016})},\ \Eprint {https://arxiv.org/abs/1606.07437} {arXiv:1606.07437 [astro-ph.HE]} \BibitemShut {NoStop}%
\bibitem [{\citenamefont {Franciolini}\ \emph {et~al.}(2022{\natexlab{a}})\citenamefont {Franciolini}, \citenamefont {Kritos}, \citenamefont {Berti},\ and\ \citenamefont {Silk}}]{Kritos}%
  \BibitemOpen
  \bibfield  {author} {\bibinfo {author} {\bibfnamefont {G.}~\bibnamefont {Franciolini}}, \bibinfo {author} {\bibfnamefont {K.}~\bibnamefont {Kritos}}, \bibinfo {author} {\bibfnamefont {E.}~\bibnamefont {Berti}},\ and\ \bibinfo {author} {\bibfnamefont {J.}~\bibnamefont {Silk}},\ }\bibfield  {title} {\bibinfo {title} {{Primordial black hole mergers from three-body interactions}},\ }\href {https://doi.org/10.1103/PhysRevD.106.083529} {\bibfield  {journal} {\bibinfo  {journal} {Phys. Rev. D}\ }\textbf {\bibinfo {volume} {106}},\ \bibinfo {pages} {083529} (\bibinfo {year} {2022}{\natexlab{a}})},\ \Eprint {https://arxiv.org/abs/2205.15340} {arXiv:2205.15340 [astro-ph.CO]} \BibitemShut {NoStop}%
\bibitem [{\citenamefont {Navarro}\ \emph {et~al.}(1996)\citenamefont {Navarro}, \citenamefont {Frenk},\ and\ \citenamefont {White}}]{Navarro:1995iw}%
  \BibitemOpen
  \bibfield  {author} {\bibinfo {author} {\bibfnamefont {J.~F.}\ \bibnamefont {Navarro}}, \bibinfo {author} {\bibfnamefont {C.~S.}\ \bibnamefont {Frenk}},\ and\ \bibinfo {author} {\bibfnamefont {S.~D.~M.}\ \bibnamefont {White}},\ }\bibfield  {title} {\bibinfo {title} {{The Structure of cold dark matter halos}},\ }\href {https://doi.org/10.1086/177173} {\bibfield  {journal} {\bibinfo  {journal} {Astrophys. J.}\ }\textbf {\bibinfo {volume} {462}},\ \bibinfo {pages} {563} (\bibinfo {year} {1996})},\ \Eprint {https://arxiv.org/abs/astro-ph/9508025} {arXiv:astro-ph/9508025} \BibitemShut {NoStop}%
\bibitem [{\citenamefont {Prada}\ \emph {et~al.}(2012)\citenamefont {Prada}, \citenamefont {Klypin}, \citenamefont {Cuesta}, \citenamefont {Betancort-Rijo},\ and\ \citenamefont {Primack}}]{Prada12}%
  \BibitemOpen
  \bibfield  {author} {\bibinfo {author} {\bibfnamefont {F.}~\bibnamefont {Prada}}, \bibinfo {author} {\bibfnamefont {A.~A.}\ \bibnamefont {Klypin}}, \bibinfo {author} {\bibfnamefont {A.~J.}\ \bibnamefont {Cuesta}}, \bibinfo {author} {\bibfnamefont {J.~E.}\ \bibnamefont {Betancort-Rijo}},\ and\ \bibinfo {author} {\bibfnamefont {J.}~\bibnamefont {Primack}},\ }\bibfield  {title} {\bibinfo {title} {{Halo concentrations in the standard LCDM cosmology}},\ }\href {https://doi.org/10.1111/j.1365-2966.2012.21007.x} {\bibfield  {journal} {\bibinfo  {journal} {Mon. Not. Roy. Astron. Soc.}\ }\textbf {\bibinfo {volume} {423}},\ \bibinfo {pages} {3018} (\bibinfo {year} {2012})},\ \Eprint {https://arxiv.org/abs/1104.5130} {arXiv:1104.5130 [astro-ph.CO]} \BibitemShut {NoStop}%
\bibitem [{\citenamefont {Ludlow}\ \emph {et~al.}(2016)\citenamefont {Ludlow}, \citenamefont {Bose}, \citenamefont {Angulo}, \citenamefont {Wang}, \citenamefont {Hellwing}, \citenamefont {Navarro}, \citenamefont {Cole},\ and\ \citenamefont {Frenk}}]{Ludlow:2016ifl}%
  \BibitemOpen
  \bibfield  {author} {\bibinfo {author} {\bibfnamefont {A.~D.}\ \bibnamefont {Ludlow}}, \bibinfo {author} {\bibfnamefont {S.}~\bibnamefont {Bose}}, \bibinfo {author} {\bibfnamefont {R.~E.}\ \bibnamefont {Angulo}}, \bibinfo {author} {\bibfnamefont {L.}~\bibnamefont {Wang}}, \bibinfo {author} {\bibfnamefont {W.~A.}\ \bibnamefont {Hellwing}}, \bibinfo {author} {\bibfnamefont {J.~F.}\ \bibnamefont {Navarro}}, \bibinfo {author} {\bibfnamefont {S.}~\bibnamefont {Cole}},\ and\ \bibinfo {author} {\bibfnamefont {C.~S.}\ \bibnamefont {Frenk}},\ }\bibfield  {title} {\bibinfo {title} {{The mass\textendash{}concentration\textendash{}redshift relation of cold and warm dark matter haloes}},\ }\href {https://doi.org/10.1093/mnras/stw1046} {\bibfield  {journal} {\bibinfo  {journal} {Mon. Not. Roy. Astron. Soc.}\ }\textbf {\bibinfo {volume} {460}},\ \bibinfo {pages} {1214} (\bibinfo {year} {2016})},\ \Eprint {https://arxiv.org/abs/1601.02624} {arXiv:1601.02624 [astro-ph.CO]} \BibitemShut {NoStop}%
\bibitem [{\citenamefont {{Press}}\ and\ \citenamefont {{Schechter}}(1974)}]{PS}%
  \BibitemOpen
  \bibfield  {author} {\bibinfo {author} {\bibfnamefont {W.~H.}\ \bibnamefont {{Press}}}\ and\ \bibinfo {author} {\bibfnamefont {P.}~\bibnamefont {{Schechter}}},\ }\bibfield  {title} {\bibinfo {title} {{Formation of Galaxies and Clusters of Galaxies by Self-Similar Gravitational Condensation}},\ }\href {https://doi.org/10.1086/152650} {\bibfield  {journal} {\bibinfo  {journal} {\apj}\ }\textbf {\bibinfo {volume} {187}},\ \bibinfo {pages} {425} (\bibinfo {year} {1974})}\BibitemShut {NoStop}%
\bibitem [{\citenamefont {Murray}\ \emph {et~al.}(2013)\citenamefont {Murray}, \citenamefont {Power},\ and\ \citenamefont {Robotham}}]{HMFcalc}%
  \BibitemOpen
  \bibfield  {author} {\bibinfo {author} {\bibfnamefont {S.}~\bibnamefont {Murray}}, \bibinfo {author} {\bibfnamefont {C.}~\bibnamefont {Power}},\ and\ \bibinfo {author} {\bibfnamefont {A.~S.~G.}\ \bibnamefont {Robotham}},\ }\bibfield  {title} {\bibinfo {title} {{HMFcalc: An online tool for calculating dark matter halo mass functions}},\ }\href {https://doi.org/10.1016/j.ascom.2013.11.001} {\bibfield  {journal} {\bibinfo  {journal} {Astron. Comput.}\ }\textbf {\bibinfo {volume} {3-4}},\ \bibinfo {pages} {23} (\bibinfo {year} {2013})},\ \Eprint {https://arxiv.org/abs/1306.6721} {arXiv:1306.6721 [astro-ph.CO]} \BibitemShut {NoStop}%
\bibitem [{\citenamefont {Franciolini}\ \emph {et~al.}(2022{\natexlab{b}})\citenamefont {Franciolini}, \citenamefont {Kritos}, \citenamefont {Berti},\ and\ \citenamefont {Silk}}]{Franciolini:2022ewd}%
  \BibitemOpen
  \bibfield  {author} {\bibinfo {author} {\bibfnamefont {G.}~\bibnamefont {Franciolini}}, \bibinfo {author} {\bibfnamefont {K.}~\bibnamefont {Kritos}}, \bibinfo {author} {\bibfnamefont {E.}~\bibnamefont {Berti}},\ and\ \bibinfo {author} {\bibfnamefont {J.}~\bibnamefont {Silk}},\ }\bibfield  {title} {\bibinfo {title} {{Primordial black hole mergers from three-body interactions}},\ }\href {https://doi.org/10.1103/PhysRevD.106.083529} {\bibfield  {journal} {\bibinfo  {journal} {Phys. Rev. D}\ }\textbf {\bibinfo {volume} {106}},\ \bibinfo {pages} {083529} (\bibinfo {year} {2022}{\natexlab{b}})},\ \Eprint {https://arxiv.org/abs/2205.15340} {arXiv:2205.15340 [astro-ph.CO]} \BibitemShut {NoStop}%
\bibitem [{\citenamefont {El~Bouhaddouti}\ \emph {et~al.}(2024)\citenamefont {El~Bouhaddouti}, \citenamefont {Aljaf},\ and\ \citenamefont {Cholis}}]{BAC:2024}%
  \BibitemOpen
  \bibfield  {author} {\bibinfo {author} {\bibfnamefont {M.}~\bibnamefont {El~Bouhaddouti}}, \bibinfo {author} {\bibfnamefont {M.}~\bibnamefont {Aljaf}},\ and\ \bibinfo {author} {\bibfnamefont {I.}~\bibnamefont {Cholis}},\ }\bibfield  {title} {\bibinfo {title} {{Consevative limits on primordial black holes from the LIGO-Virgo-KAGRA observations}},\ }\href@noop {} {\  (\bibinfo {year} {2024})},\ \Eprint {https://arxiv.org/abs/in preparation} {arXiv:in preparation} \BibitemShut {NoStop}%
\bibitem [{\citenamefont {Kavanagh}\ \emph {et~al.}(2018{\natexlab{a}})\citenamefont {Kavanagh}, \citenamefont {Gaggero},\ and\ \citenamefont {Bertone}}]{Kavanagh:2018ggo}%
  \BibitemOpen
  \bibfield  {author} {\bibinfo {author} {\bibfnamefont {B.~J.}\ \bibnamefont {Kavanagh}}, \bibinfo {author} {\bibfnamefont {D.}~\bibnamefont {Gaggero}},\ and\ \bibinfo {author} {\bibfnamefont {G.}~\bibnamefont {Bertone}},\ }\bibfield  {title} {\bibinfo {title} {{Merger rate of a subdominant population of primordial black holes}},\ }\href {https://doi.org/10.1103/PhysRevD.98.023536} {\bibfield  {journal} {\bibinfo  {journal} {Phys. Rev. D}\ }\textbf {\bibinfo {volume} {98}},\ \bibinfo {pages} {023536} (\bibinfo {year} {2018}{\natexlab{a}})},\ \Eprint {https://arxiv.org/abs/1805.09034} {arXiv:1805.09034 [astro-ph.CO]} \BibitemShut {NoStop}%
\bibitem [{\citenamefont {Peters}(1964)}]{Peters}%
  \BibitemOpen
  \bibfield  {author} {\bibinfo {author} {\bibfnamefont {P.~C.}\ \bibnamefont {Peters}},\ }\bibfield  {title} {\bibinfo {title} {Gravitational radiation and the motion of two point masses},\ }\href {https://doi.org/10.1103/PhysRev.136.B1224} {\bibfield  {journal} {\bibinfo  {journal} {Phys. Rev.}\ }\textbf {\bibinfo {volume} {136}},\ \bibinfo {pages} {B1224} (\bibinfo {year} {1964})}\BibitemShut {NoStop}%
\bibitem [{\citenamefont {Quinlan}(1996)}]{H_rate}%
  \BibitemOpen
  \bibfield  {author} {\bibinfo {author} {\bibfnamefont {G.~D.}\ \bibnamefont {Quinlan}},\ }\bibfield  {title} {\bibinfo {title} {{The dynamical evolution of massive black hole binaries - I. hardening in a fixed stellar background}},\ }\href {https://doi.org/10.1016/S1384-1076(96)00003-6} {\bibfield  {journal} {\bibinfo  {journal} {New Astron.}\ }\textbf {\bibinfo {volume} {1}},\ \bibinfo {pages} {35} (\bibinfo {year} {1996})},\ \Eprint {https://arxiv.org/abs/astro-ph/9601092} {arXiv:astro-ph/9601092} \BibitemShut {NoStop}%
\bibitem [{\citenamefont {Sesana}\ \emph {et~al.}(2006)\citenamefont {Sesana}, \citenamefont {Haardt},\ and\ \citenamefont {Madau}}]{H_approx}%
  \BibitemOpen
  \bibfield  {author} {\bibinfo {author} {\bibfnamefont {A.}~\bibnamefont {Sesana}}, \bibinfo {author} {\bibfnamefont {F.}~\bibnamefont {Haardt}},\ and\ \bibinfo {author} {\bibfnamefont {P.}~\bibnamefont {Madau}},\ }\bibfield  {title} {\bibinfo {title} {{Interaction of massive black hole binaries with their stellar environment. 1. Ejection of hypervelocity stars}},\ }\href {https://doi.org/10.1086/507596} {\bibfield  {journal} {\bibinfo  {journal} {Astrophys. J.}\ }\textbf {\bibinfo {volume} {651}},\ \bibinfo {pages} {392} (\bibinfo {year} {2006})},\ \Eprint {https://arxiv.org/abs/astro-ph/0604299} {arXiv:astro-ph/0604299} \BibitemShut {NoStop}%
\bibitem [{\citenamefont {Correa}\ \emph {et~al.}(2015)\citenamefont {Correa}, \citenamefont {Wyithe}, \citenamefont {Schaye},\ and\ \citenamefont {Duffy}}]{MAH}%
  \BibitemOpen
  \bibfield  {author} {\bibinfo {author} {\bibfnamefont {C.~A.}\ \bibnamefont {Correa}}, \bibinfo {author} {\bibfnamefont {J.~S.~B.}\ \bibnamefont {Wyithe}}, \bibinfo {author} {\bibfnamefont {J.}~\bibnamefont {Schaye}},\ and\ \bibinfo {author} {\bibfnamefont {A.~R.}\ \bibnamefont {Duffy}},\ }\bibfield  {title} {\bibinfo {title} {{The accretion history of dark matter haloes \textendash{} II. The connections with the mass power spectrum and the density profile}},\ }\href {https://doi.org/10.1093/mnras/stv697} {\bibfield  {journal} {\bibinfo  {journal} {Mon. Not. Roy. Astron. Soc.}\ }\textbf {\bibinfo {volume} {450}},\ \bibinfo {pages} {1521} (\bibinfo {year} {2015})},\ \Eprint {https://arxiv.org/abs/1501.04382} {arXiv:1501.04382 [astro-ph.CO]} \BibitemShut {NoStop}%
\bibitem [{\citenamefont {Heggie}(1975)}]{Heggie}%
  \BibitemOpen
  \bibfield  {author} {\bibinfo {author} {\bibfnamefont {D.~C.}\ \bibnamefont {Heggie}},\ }\bibfield  {title} {\bibinfo {title} {{Binary Evolution in Stellar Dynamics}},\ }\href {https://doi.org/10.1093/mnras/173.3.729} {\bibfield  {journal} {\bibinfo  {journal} {Mon. Not. Roy. Astron. Soc.}\ }\textbf {\bibinfo {volume} {173}},\ \bibinfo {pages} {729} (\bibinfo {year} {1975})}\BibitemShut {NoStop}%
\bibitem [{\citenamefont {Kavanagh}\ \emph {et~al.}(2018{\natexlab{b}})\citenamefont {Kavanagh}, \citenamefont {Gaggero},\ and\ \citenamefont {Bertone}}]{PBBH2}%
  \BibitemOpen
  \bibfield  {author} {\bibinfo {author} {\bibfnamefont {B.~J.}\ \bibnamefont {Kavanagh}}, \bibinfo {author} {\bibfnamefont {D.}~\bibnamefont {Gaggero}},\ and\ \bibinfo {author} {\bibfnamefont {G.}~\bibnamefont {Bertone}},\ }\bibfield  {title} {\bibinfo {title} {{Merger rate of a subdominant population of primordial black holes}},\ }\href {https://doi.org/10.1103/PhysRevD.98.023536} {\bibfield  {journal} {\bibinfo  {journal} {Phys. Rev. D}\ }\textbf {\bibinfo {volume} {98}},\ \bibinfo {pages} {023536} (\bibinfo {year} {2018}{\natexlab{b}})},\ \Eprint {https://arxiv.org/abs/1805.09034} {arXiv:1805.09034 [astro-ph.CO]} \BibitemShut {NoStop}%
\bibitem [{\citenamefont {Liu}\ \emph {et~al.}(2019)\citenamefont {Liu}, \citenamefont {Guo},\ and\ \citenamefont {Cai}}]{PBBH3}%
  \BibitemOpen
  \bibfield  {author} {\bibinfo {author} {\bibfnamefont {L.}~\bibnamefont {Liu}}, \bibinfo {author} {\bibfnamefont {Z.-K.}\ \bibnamefont {Guo}},\ and\ \bibinfo {author} {\bibfnamefont {R.-G.}\ \bibnamefont {Cai}},\ }\bibfield  {title} {\bibinfo {title} {{Effects of the surrounding primordial black holes on the merger rate of primordial black hole binaries}},\ }\href {https://doi.org/10.1103/PhysRevD.99.063523} {\bibfield  {journal} {\bibinfo  {journal} {Phys. Rev. D}\ }\textbf {\bibinfo {volume} {99}},\ \bibinfo {pages} {063523} (\bibinfo {year} {2019})},\ \Eprint {https://arxiv.org/abs/1812.05376} {arXiv:1812.05376 [astro-ph.CO]} \BibitemShut {NoStop}%
\bibitem [{\citenamefont {Jenkins}\ \emph {et~al.}(2001)\citenamefont {Jenkins}, \citenamefont {Frenk}, \citenamefont {White}, \citenamefont {Colberg}, \citenamefont {Cole}, \citenamefont {Evrard}, \citenamefont {Couchman},\ and\ \citenamefont {Yoshida}}]{Jenkins}%
  \BibitemOpen
  \bibfield  {author} {\bibinfo {author} {\bibfnamefont {A.}~\bibnamefont {Jenkins}}, \bibinfo {author} {\bibfnamefont {C.~S.}\ \bibnamefont {Frenk}}, \bibinfo {author} {\bibfnamefont {S.~D.~M.}\ \bibnamefont {White}}, \bibinfo {author} {\bibfnamefont {J.~M.}\ \bibnamefont {Colberg}}, \bibinfo {author} {\bibfnamefont {S.}~\bibnamefont {Cole}}, \bibinfo {author} {\bibfnamefont {A.~E.}\ \bibnamefont {Evrard}}, \bibinfo {author} {\bibfnamefont {H.~M.~P.}\ \bibnamefont {Couchman}},\ and\ \bibinfo {author} {\bibfnamefont {N.}~\bibnamefont {Yoshida}},\ }\bibfield  {title} {\bibinfo {title} {{The Mass function of dark matter halos}},\ }\href {https://doi.org/10.1046/j.1365-8711.2001.04029.x} {\bibfield  {journal} {\bibinfo  {journal} {Mon. Not. Roy. Astron. Soc.}\ }\textbf {\bibinfo {volume} {321}},\ \bibinfo {pages} {372} (\bibinfo {year} {2001})},\ \Eprint {https://arxiv.org/abs/astro-ph/0005260} {arXiv:astro-ph/0005260} \BibitemShut {NoStop}%
\bibitem [{\citenamefont {Martinelli}\ \emph {et~al.}(2022)\citenamefont {Martinelli}, \citenamefont {Scarcella}, \citenamefont {Hogg}, \citenamefont {Kavanagh}, \citenamefont {Gaggero},\ and\ \citenamefont {Fleury}}]{Martinelli:2022elq}%
  \BibitemOpen
  \bibfield  {author} {\bibinfo {author} {\bibfnamefont {M.}~\bibnamefont {Martinelli}}, \bibinfo {author} {\bibfnamefont {F.}~\bibnamefont {Scarcella}}, \bibinfo {author} {\bibfnamefont {N.~B.}\ \bibnamefont {Hogg}}, \bibinfo {author} {\bibfnamefont {B.~J.}\ \bibnamefont {Kavanagh}}, \bibinfo {author} {\bibfnamefont {D.}~\bibnamefont {Gaggero}},\ and\ \bibinfo {author} {\bibfnamefont {P.}~\bibnamefont {Fleury}},\ }\bibfield  {title} {\bibinfo {title} {{Dancing in the dark: detecting a population of distant primordial black holes}},\ }\href {https://doi.org/10.1088/1475-7516/2022/08/006} {\bibfield  {journal} {\bibinfo  {journal} {JCAP}\ }\textbf {\bibinfo {volume} {08}}\bibfield  {number} {\bibinfo  {number} { (08)},\ \bibinfo {pages} {006}},\ }\Eprint {https://arxiv.org/abs/2205.02639} {arXiv:2205.02639 [astro-ph.CO]} \BibitemShut {NoStop}%
\bibitem [{\citenamefont {Blaineau}\ \emph {et~al.}(2022)\citenamefont {Blaineau} \emph {et~al.}}]{Blaineau:2022nhy}%
  \BibitemOpen
  \bibfield  {author} {\bibinfo {author} {\bibfnamefont {T.}~\bibnamefont {Blaineau}} \emph {et~al.},\ }\bibfield  {title} {\bibinfo {title} {{New limits from microlensing on Galactic black holes in the mass range 10 M\ensuremath{\odot} \ensuremath{<} M \ensuremath{<} 1000 M\ensuremath{\odot}}},\ }\href {https://doi.org/10.1051/0004-6361/202243430} {\bibfield  {journal} {\bibinfo  {journal} {Astron. Astrophys.}\ }\textbf {\bibinfo {volume} {664}},\ \bibinfo {pages} {A106} (\bibinfo {year} {2022})},\ \Eprint {https://arxiv.org/abs/2202.13819} {arXiv:2202.13819 [astro-ph.GA]} \BibitemShut {NoStop}%
\bibitem [{\citenamefont {Mroz}\ \emph {et~al.}(2024)\citenamefont {Mroz} \emph {et~al.}}]{Mroz:2024mse}%
  \BibitemOpen
  \bibfield  {author} {\bibinfo {author} {\bibfnamefont {P.}~\bibnamefont {Mroz}} \emph {et~al.},\ }\bibfield  {title} {\bibinfo {title} {{No massive black holes in the Milky Way halo}}\ }\href {https://doi.org/10.1038/s41586-024-07704-6} {10.1038/s41586-024-07704-6} (\bibinfo {year} {2024}),\ \Eprint {https://arxiv.org/abs/2403.02386} {arXiv:2403.02386 [astro-ph.GA]} \BibitemShut {NoStop}%
\bibitem [{\citenamefont {Garcia-Bellido}\ and\ \citenamefont {Hawkins}(2024)}]{Garcia-Bellido:2024yaz}%
  \BibitemOpen
  \bibfield  {author} {\bibinfo {author} {\bibfnamefont {J.}~\bibnamefont {Garcia-Bellido}}\ and\ \bibinfo {author} {\bibfnamefont {M.}~\bibnamefont {Hawkins}},\ }\bibfield  {title} {\bibinfo {title} {{Reanalysis of the MACHO constraints on PBH in the light of Gaia DR3 data}},\ }\href@noop {} {\  (\bibinfo {year} {2024})},\ \Eprint {https://arxiv.org/abs/2402.00212} {arXiv:2402.00212 [astro-ph.GA]} \BibitemShut {NoStop}%
\bibitem [{\citenamefont {Hild}\ \emph {et~al.}(2011)\citenamefont {Hild} \emph {et~al.}}]{Hild:2010id}%
  \BibitemOpen
  \bibfield  {author} {\bibinfo {author} {\bibfnamefont {S.}~\bibnamefont {Hild}} \emph {et~al.},\ }\bibfield  {title} {\bibinfo {title} {{Sensitivity Studies for Third-Generation Gravitational Wave Observatories}},\ }\href {https://doi.org/10.1088/0264-9381/28/9/094013} {\bibfield  {journal} {\bibinfo  {journal} {Class. Quant. Grav.}\ }\textbf {\bibinfo {volume} {28}},\ \bibinfo {pages} {094013} (\bibinfo {year} {2011})},\ \Eprint {https://arxiv.org/abs/1012.0908} {arXiv:1012.0908 [gr-qc]} \BibitemShut {NoStop}%
\bibitem [{\citenamefont {Evans}\ \emph {et~al.}(2023)\citenamefont {Evans} \emph {et~al.}}]{Evans:2023euw}%
  \BibitemOpen
  \bibfield  {author} {\bibinfo {author} {\bibfnamefont {M.}~\bibnamefont {Evans}} \emph {et~al.},\ }\bibfield  {title} {\bibinfo {title} {{Cosmic Explorer: A Submission to the NSF MPSAC ngGW Subcommittee}},\ }\href@noop {} {\  (\bibinfo {year} {2023})},\ \Eprint {https://arxiv.org/abs/2306.13745} {arXiv:2306.13745 [astro-ph.IM]} \BibitemShut {NoStop}%
\bibitem [{\citenamefont {Gupta}\ \emph {et~al.}(2023)\citenamefont {Gupta} \emph {et~al.}}]{Gupta:2023lga}%
  \BibitemOpen
  \bibfield  {author} {\bibinfo {author} {\bibfnamefont {I.}~\bibnamefont {Gupta}} \emph {et~al.},\ }\bibfield  {title} {\bibinfo {title} {{Characterizing Gravitational Wave Detector Networks: From A$^\sharp$ to Cosmic Explorer}},\ }\href@noop {} {\  (\bibinfo {year} {2023})},\ \Eprint {https://arxiv.org/abs/2307.10421} {arXiv:2307.10421 [gr-qc]} \BibitemShut {NoStop}%
\bibitem [{\citenamefont {Chen}\ and\ \citenamefont {Hall}(2024)}]{Chen:2024dxh}%
  \BibitemOpen
  \bibfield  {author} {\bibinfo {author} {\bibfnamefont {Z.-C.}\ \bibnamefont {Chen}}\ and\ \bibinfo {author} {\bibfnamefont {A.}~\bibnamefont {Hall}},\ }\bibfield  {title} {\bibinfo {title} {{Confronting primordial black holes with LIGO-Virgo-KAGRA and the Einstein Telescope}},\ }\href@noop {} {\  (\bibinfo {year} {2024})},\ \Eprint {https://arxiv.org/abs/2402.03934} {arXiv:2402.03934 [astro-ph.CO]} \BibitemShut {NoStop}%
\bibitem [{\citenamefont {Franciolini}\ \emph {et~al.}(2022{\natexlab{c}})\citenamefont {Franciolini}, \citenamefont {Cotesta}, \citenamefont {Loutrel}, \citenamefont {Berti}, \citenamefont {Pani},\ and\ \citenamefont {Riotto}}]{PBBH4}%
  \BibitemOpen
  \bibfield  {author} {\bibinfo {author} {\bibfnamefont {G.}~\bibnamefont {Franciolini}}, \bibinfo {author} {\bibfnamefont {R.}~\bibnamefont {Cotesta}}, \bibinfo {author} {\bibfnamefont {N.}~\bibnamefont {Loutrel}}, \bibinfo {author} {\bibfnamefont {E.}~\bibnamefont {Berti}}, \bibinfo {author} {\bibfnamefont {P.}~\bibnamefont {Pani}},\ and\ \bibinfo {author} {\bibfnamefont {A.}~\bibnamefont {Riotto}},\ }\bibfield  {title} {\bibinfo {title} {{How to assess the primordial origin of single gravitational-wave events with mass, spin, eccentricity, and deformability measurements}},\ }\href {https://doi.org/10.1103/PhysRevD.105.063510} {\bibfield  {journal} {\bibinfo  {journal} {Phys. Rev. D}\ }\textbf {\bibinfo {volume} {105}},\ \bibinfo {pages} {063510} (\bibinfo {year} {2022}{\natexlab{c}})},\ \Eprint {https://arxiv.org/abs/2112.10660} {arXiv:2112.10660 [astro-ph.CO]} \BibitemShut {NoStop}%
\end{thebibliography}%
\end{document}